\documentclass{aa}

\usepackage{graphicx}
\usepackage{txfonts}
\usepackage{longtable}
\usepackage{subcaption}
\usepackage{lscape}
\usepackage{xcolor}
\usepackage{multirow}
\usepackage{array}
\usepackage{longtable}

\def\cd{\,\mathrm{c/d}}

\begin{document}

   \title{Harmonics as a Hidden Window into the Turbulent Convective Envelope of non-Blazhko RRab Stars}
\titlerunning{Harmonics in RRab Stars}

   \author{Jia-Shu Niu\inst{1,2,3}\corrauth{jsniu@sxu.edu.cn}}
   \institute{Institute of Theoretical Physics, Shanxi University, Taiyuan 030006, China;\\
        \and
        State Key Laboratory of Quantum Optics Technologies and Devices, Shanxi University, Taiyuan 030006, China;\\
        \and
             Collaborative Innovation Center of Extreme Optics, Shanxi University, Taiyuan 030006, China;\\
             }

   \date{Received XX XX, 2026; accepted XX XX, 2026}

\abstract
{Harmonics in pulsating stars have traditionally been regarded as mere Fourier artifacts, fully determined by their parent mode. Recent observations of high-amplitude $\delta$ Scuti stars, however, have revealed the existence of disharmonized harmonics, which exhibit amplitude and frequency variations uncorrelated with their parent pulsation mode. Here we test the universality of this phenomenon by examining a class of stars long considered pulsationally stable: non-Blazhko RRab stars. Using short-cadence {\it Kepler} photometric data, we not only identify a distinct hump structure formed by harmonics of the primary pulsation mode in their frequency spectra, but also find significant amplitude and frequency variations associated with the harmonics around the onset and in the tail of the hump structure. These findings are consistent with the interpretation that the hump structure originates from the turbulent convective envelope of the star. Interestingly, several important phenomena can be understood within this framework as a working hypothesis. Thus, the hump structure of harmonics and their intrinsic variability could constitute a hidden window into the stellar convective envelope, potentially tracing energy injection at the convection-pulsation interaction and stochastic perturbations from turbulent convection -- although detailed modeling is required to confirm this interpretation.}


   \maketitle
\nolinenumbers

\section{Introduction}
\label{sec:intro}

RR Lyrae stars are low-mass, large-amplitude pulsating stars on the horizontal branch, and they are fundamental to establishing the cosmic distance ladder within the Milky Way and beyond \citep{Kurtz2022}. Their pulsations, primarily classified into fundamental (RRab), first overtone (RRc), and fundamental-first overtone hybrid (RRd) modes \citep{Plachy2021}, are self-excited by the classical $\kappa$ mechanism operating in the He~II ionization zone \citep{Baker1962, Cox1963}, with the outer He~I/H ionization region playing a secondary, supportive role \citep{Bono1994}. Crucially, the convective envelope, which spatially overlaps with this outer ionization zone, is not a passive bystander; it is essential for replicating observed light curves \citep{Unno1967, Baker1979} and is theorized to be the seat of vigorous convection-pulsation interaction \citep{Houdek2015}.

A ubiquitous yet often overlooked feature in the light curves of these stars is the presence of harmonics. Detected across a range of variable stars, from Cepheids \citep{Rathour2021} to RR Lyrae \citep{Kurtz2016} and high-amplitude $\delta$ Scuti stars \citep{Niu2022}, harmonics have historically been dismissed as mere byproducts of the primary, sinusoidal pulsation---simple Fourier artifacts with no independent physical significance. This view has effectively excluded them from asteroseismic modeling.

This conventional wisdom has recently been challenged. In high-amplitude $\delta$ Scuti stars, a subset of harmonics has been found to behave as nearly independent entities, exhibiting amplitude and frequency variations that are completely uncorrelated with their parent mode \citep{Niu2023, Niu2024, Xue2024}. Termed `disharmonized harmonics,' they suggest a more complex reality: harmonics are not totally determined by their parent pulsation modes \citep{Niu2022, Niu2024, Niu2025}, and their very generation may be a fingerprint of distinct nonlinear processes \citep{Xue2024}. This discovery has cracked the edifice of the traditional view, but it has also left a critical question: are disharmonized harmonics a peculiarity of high-amplitude $\delta$ Scuti stars, or do they represent a universal phenomenon present across many kinds of pulsating stars? If universal, their implications would be profound. They would not only necessitate a revision of the textbook understanding of harmonics but could also transform them into a powerful, high-sensitivity probe of the most inaccessible regions of stellar interiors---particularly the outer convective envelope and its dynamic interaction with the pulsation.

Here, we test this hypothesis by turning to a seemingly unlikely candidate: the non-Blazhko RRab stars, a class of stars presumed to be pulsationally stable, exhibiting none of the quasi-periodic amplitude or phase modulations that have long puzzled astrophysicists. This makes them an ideal testbed: if disharmonized harmonics are a genuine and common physical phenomenon, they should manifest even here, in the absence of large-scale modulation.

\section{Methods}
\label{sec:methods}

The short-cadence (SC) photometric data from the \textit{Kepler} space telescope \citep{Nemec2011, Plachy2021} provide an excellent opportunity to perform this test. 
In this work, we select the non-Blazhko RRab stars that were monitored by {\it Kepler}'s short-cadence exposure mode and have a data time span of more than 90 days (see Table \ref{tab:kepler_RRs}). For each star, a complete quarter is used to minimize the potential impact of data artifacts on the results.

\begin{table}[htbp]
  \centering
  \caption{RRab stars studied in this work.}
  \label{tab:kepler_RRs}
  \begin{tabular}{l|cc}
    \hline
    \hline
    Variable name & SC quarter used & Data span (days) \\
    \hline
    AW Dra & Q5 & 94.7 \\
    FN Lyr & Q5 & 94.7 \\
    KIC 7030715 & Q9 & 97.4 \\
    KIC 9717032 & Q11 & 97.1 \\
    NQ Lyr & Q10 & 93.4 \\
    V346 Lyr & Q10 & 93.4 \\
    V349 Lyr & Q9 & 97.4 \\
    V368 Lyr & Q10 & 93.4 \\
    V715 Cyg & Q9 & 97.4 \\
    V782 Cyg & Q9 & 97.4 \\
    V784 Cyg & Q13 & 90.3 \\
    V894 Lyr & Q9 & 97.4 \\
    V1107 Cyg & Q9 & 97.4 \\
    V2470 Cyg & Q10 & 93.4 \\
    \hline
  \end{tabular}
\end{table}

For each star, the publicly available Pre-search Data Conditioning (PDC) light curves \citep{Kepler01, Kepler02} were retrieved from the Mikulski Archive for Space Telescopes (MAST)\footnote{\url{http://archive.stsci.edu/kepler}} using the {\tt Lightkurve} module \citep{lightkurve}. These data are provided in barycentric Julian date (BJD), PDCSAP flux, and PDCSAP flux error. After converting the normalized fluxes to magnitudes using the \textit{Kepler} magnitudes and removing long-term trends in each sub-quarter, we obtained the light curves for the subsequent analysis.

To extract the amplitudes and frequencies of the harmonic pulsation modes, we performed a standard pre-whitening procedure (see Appendix \ref{app:methods01} for more details), until the signal-to-noise ratio (S/N) was less than 8.0 (see Table \ref{tab:harmonics} for AW Dra as an example).

To investigate the temporal variations in the amplitudes and frequencies of the harmonics, we employed a short-time Fourier transform (STFT) approach, following the methodology established in our previous work \citep{Niu2023, Niu2024, Xue2024, Niu2025}. 
In this work, a sliding window of 15 days in length was advanced across the entire dataset in increments of 1.5 days for all the stars (see Appendix \ref{app:methods02} for more details). 
The STFT process covered all the harmonics except those with an S/N below 8.0 in any 15-day window, a threshold that depends sensitively on the intrinsic fluctuations of the harmonics themselves (see Figs. \ref{fig:var_amp_freq01}, \ref{fig:var_amp_freq02}, and \ref{fig:var_amp_freq03} for AW Dra as an example).

Furthermore, because a STFT with strong overlap is known to introduce apparent amplitude and frequency variability even for intrinsically stable signals \citep{Balona2014}, we used the extracted harmonics (for AW Dra, see Table \ref{tab:harmonics}) from the standard pre-whitening procedure to produce artificial light curves, and performed the same STFT process to obtain the amplitude and frequency variations. These were used as a baseline to assess the significance of the observed variations.

Based on the STFT results, a time series of amplitude and frequency values for the harmonics of each star provides two important metrics to quantify the magnitude of variation with harmonic order: the relative amplitude variation ($\Delta A/\bar{A} \equiv (A_\mathrm{max} - A_\mathrm{min})/\bar{A}$) and the absolute frequency variation ($\Delta f \equiv f_\mathrm{max} - f_\mathrm{min}$) over the full dataset, where $A_\mathrm{max}$ and $A_\mathrm{min}$ are the maximum and minimum amplitude values from the STFT, $\bar{A}$ is the mean amplitude over the full dataset, and $f_\mathrm{max}$ and $f_\mathrm{min}$ are the maximum and minimum frequency values.

\section{Results and Discussions}
\label{sec:results}
\begin{figure*}[htbp!]
  \centering
  \includegraphics[width=0.46\textwidth]{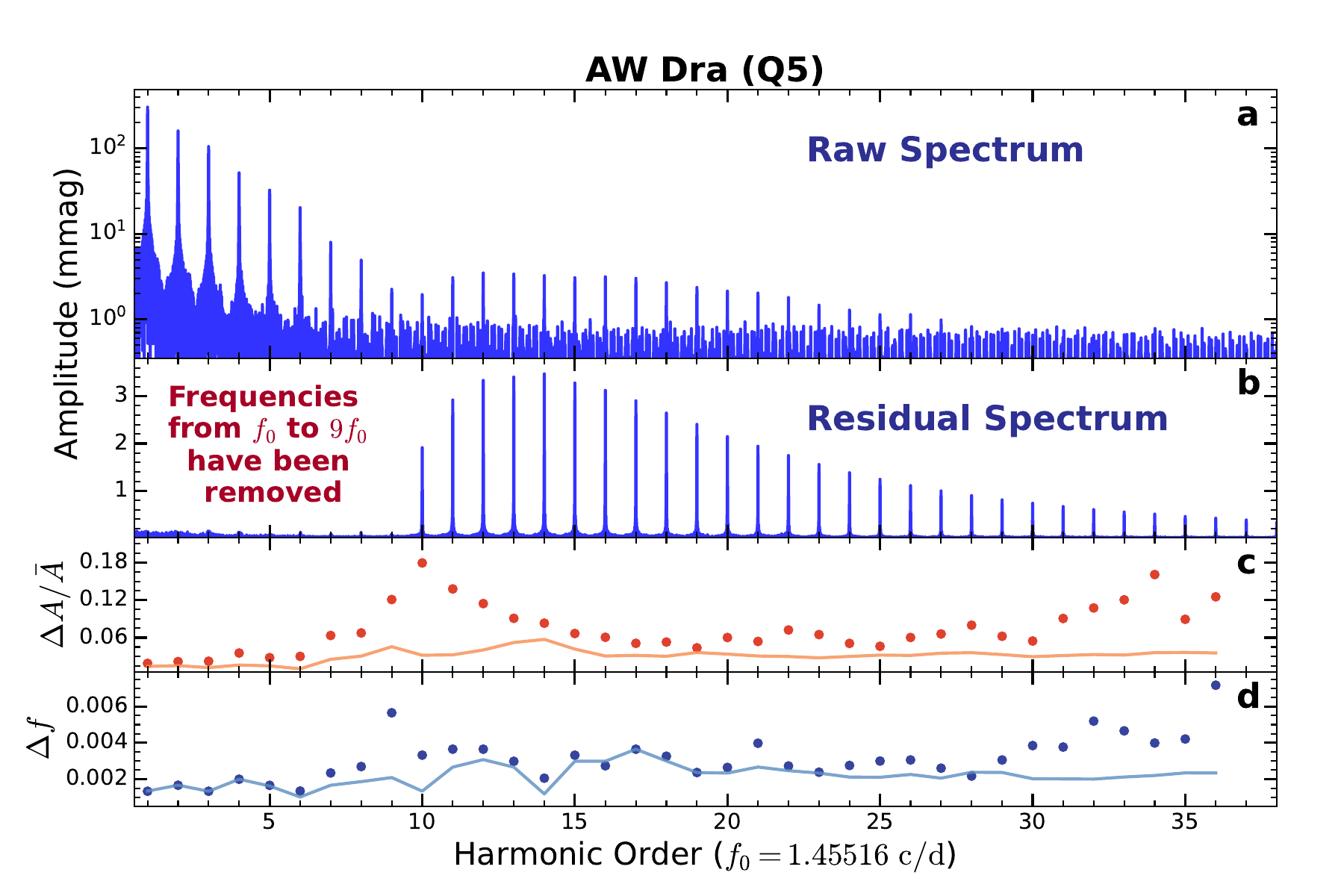}  
  \includegraphics[width=0.46\textwidth]{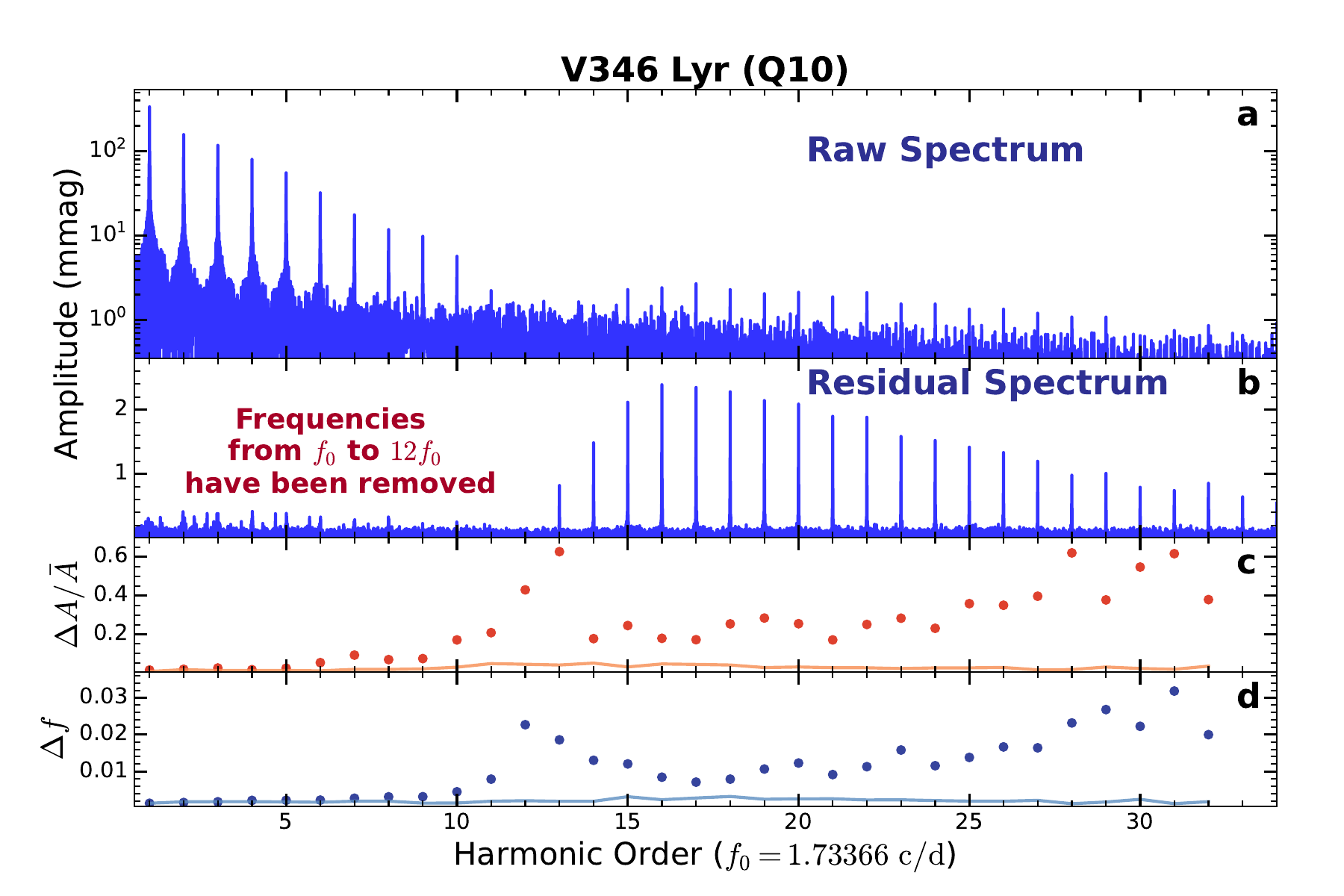}
  \includegraphics[width=0.46\textwidth]{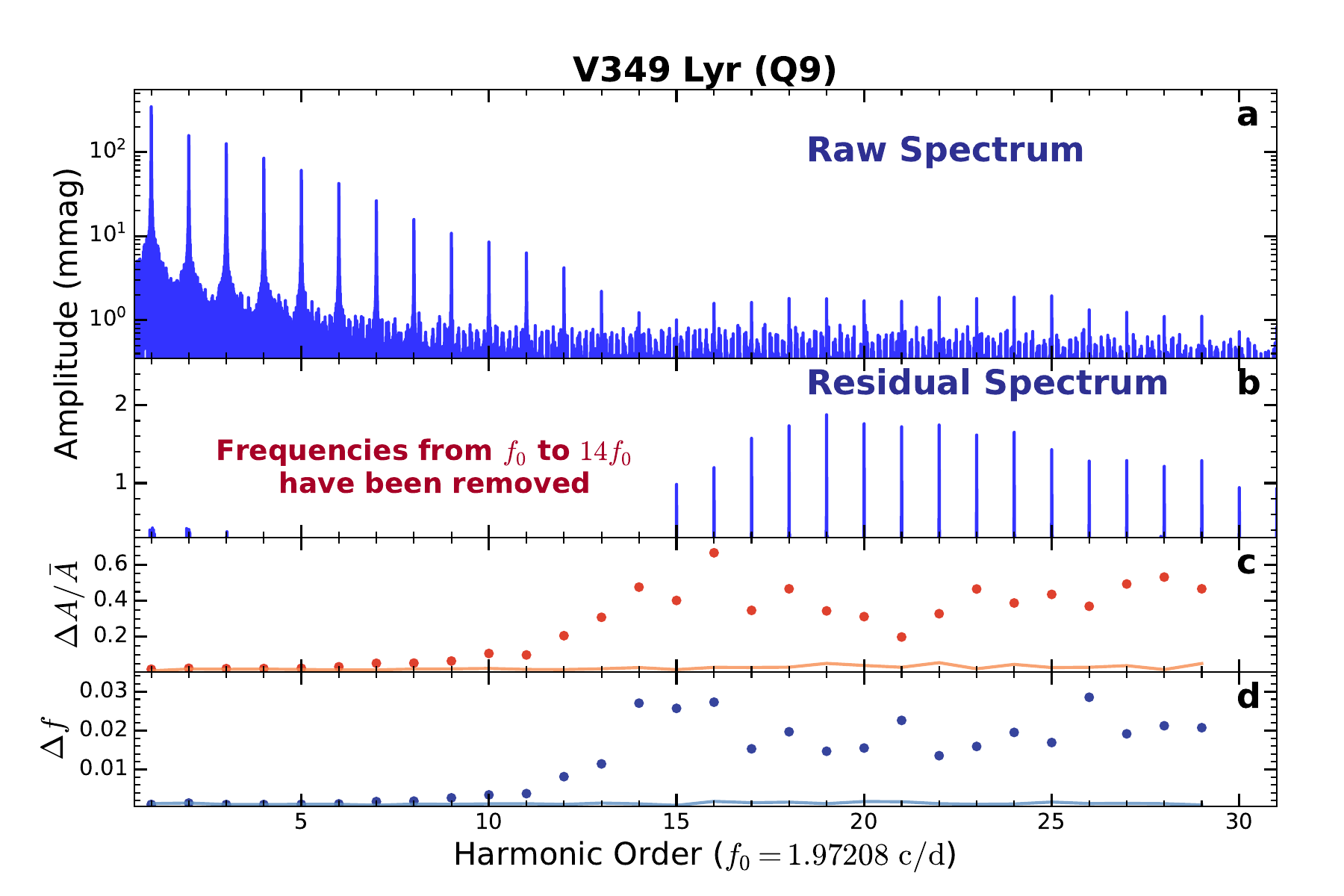}
  \includegraphics[width=0.46\textwidth]{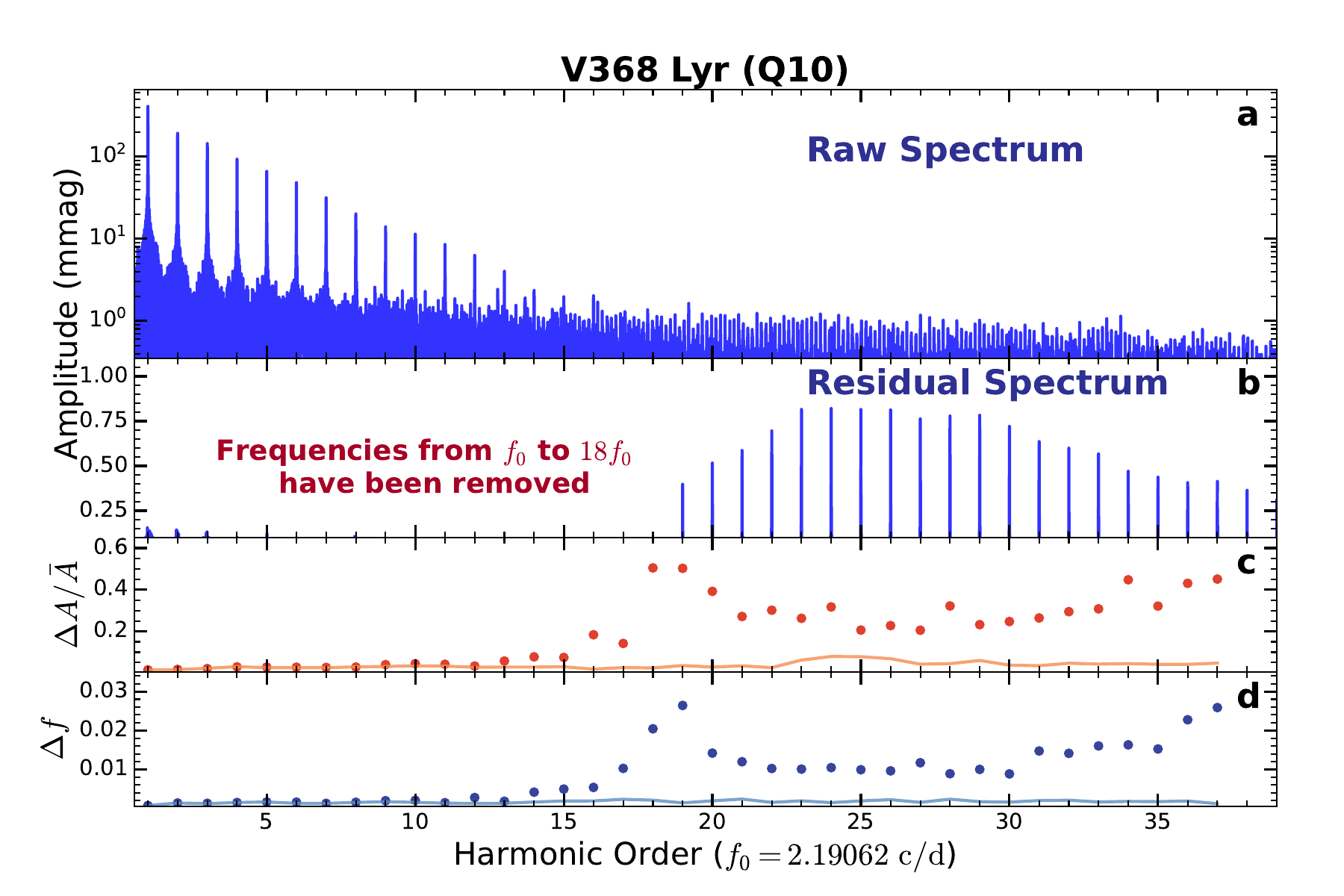}
  \includegraphics[width=0.46\textwidth]{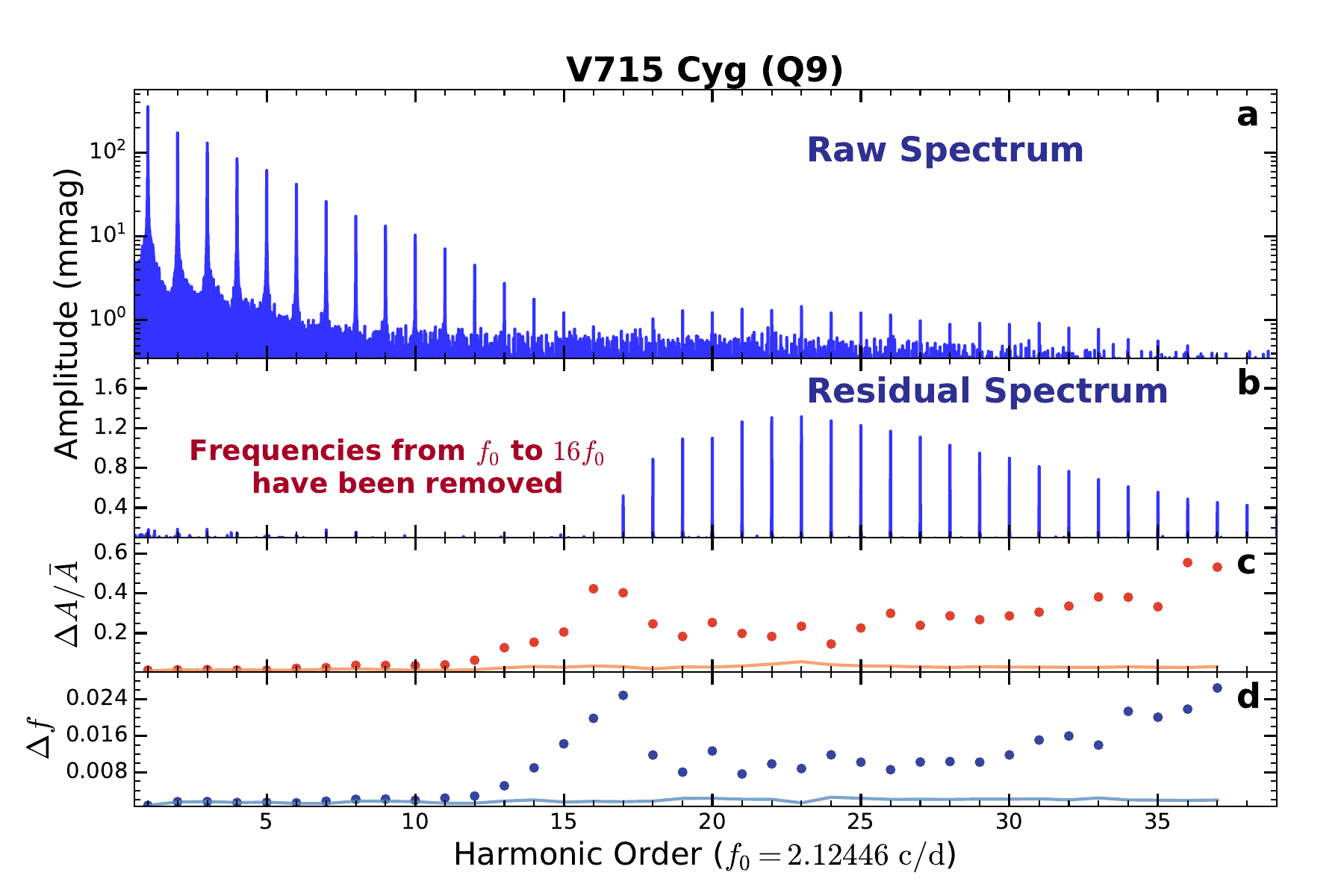}
  \includegraphics[width=0.46\textwidth]{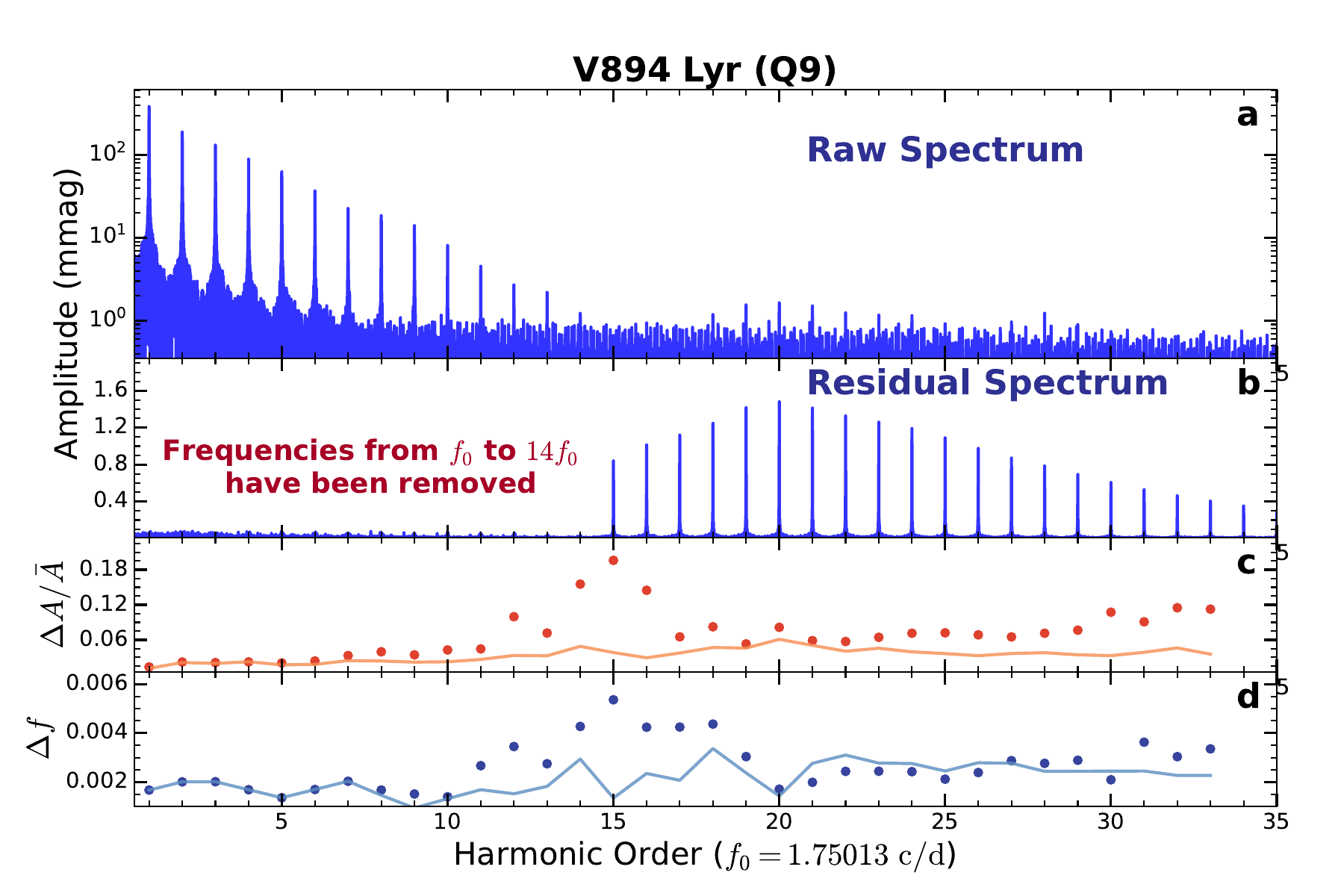}
  \caption{The harmonics in the frequency spectra and the magnitude of their amplitude and frequency variations (significant samples, whose harmonic variations are significant both around the onset and in the tail of the hump structure). (a) The raw frequency spectrum. (b) The residual spectrum after pre-whitening the exponentially decaying harmonics, highlighting the distinct hump structure at higher orders. (c) The relative amplitude variation ($\Delta A/\bar{A}$) for each harmonic over more than 90 days. (d) The absolute frequency variation ($\Delta f$) for each harmonic over more than 90 days. The light red and blue lines in panels (c) and (d) represent $\Delta A/\bar{A}$ and $\Delta f$ obtained from the artificial data, serving as baselines to demonstrate the significance of the real variations.}
\label{fig:spec_res}
\end{figure*}

Fig.~\ref{fig:spec_res} presents the distribution of harmonics and the magnitude of their amplitude and frequency variations for several significant samples (whose harmonic variations are significant both around the onset and in the tail of the hump structure), all of which display relatively large amplitude and frequency variations. Additional samples can be found in Appendix~\ref{app:add_results}, Fig.~\ref{fig:spec_res_add}.

\subsection{A Spectral Anomaly: The Harmonic Hump}
\label{sec:anomaly}

We begin with a feature that was noted and systematically analyzed by \citet{Benko2016} using CoRoT photometry of several RR Lyrae stars. The frequency spectra of these canonical non-Blazhko stars reveal a striking anomaly (Figs.~\ref{fig:spec_res} and \ref{fig:spec_res_add}). Instead of the monotonic, exponential decay of amplitude with harmonic order that might be naively expected from a simple nonlinear distortion, their harmonic sequence tells a different story. After an initial exponential decrease, the amplitudes unexpectedly surge, forming a pronounced, asymmetric hump, which reaches a peak and extends in a long, slowly decaying tail. 
Contrary to the previous description of the decline after the hump \citep{Benko2016}, we find that it decays more slowly than an exponential decline (possibly following a power law), which warrants further quantitative investigation.

This spectral morphology is not what one would expect from a pulsation wave simply propagating through a passive medium. Harmonics are generated by the nonlinear response of the stellar material. The initial exponential decay likely represents the dissipation of the pulsation energy as the wave travels from its dominant driving zone in the deep He~II ionization region outward (dominated by radiative damping). The subsequent resurgence of amplitude--the hump--is therefore a clear signature of an additional energy injection into the wave as it encounters a distinct physical environment in the outer stellar layers. The long-tailed decay beyond the hump further suggests that this outer environment has unique propagation properties, damping the wave much more slowly than the inner regions.

A natural physical correspondence presents itself. The outer stellar structure, from the base of the He~II zone to the surface, comprises the radiative zone and the He~I/H ionization zone, which might substantially overlap with the convective envelope. We propose a conceptual map: the exponentially decaying low-order harmonics trace the wave's journey through the deep interior; the rising phase of the hump marks the region where the wave first encounters and strongly interacts with the base of the convective envelope; and the long-tailed decline represents the wave's propagation through, and modification by, the turbulent convection in the stellar envelope. 
In this framework, the hump structure is consistent with a signature of convection-pulsation interaction and turbulent convection in the envelope. This interpretation is supported by the significant amplitude and frequency variations observed in the harmonics around the onset (significant samples in Fig.~\ref{fig:spec_res}) and in the tail (all samples in Figs.~\ref{fig:spec_res} and \ref{fig:spec_res_add}) of the hump structure. We emphasize that this remains a working hypothesis, pending validation through time-dependent hydrodynamic modeling.

\subsection{Hidden Variations in Harmonics}
\label{sec:variation}

For a long time, the pulsations of non-Blazhko RR Lyrae stars were considered stable. Recently, cycle-to-cycle (C2C) variations have been found in these variables based on {\it Kepler} photometric data, which are small, intrinsic, and apparently random changes in the light curve that occur between consecutive pulsation cycles \citep{Benko2019}. These variations, typically concentrated around maximum light with amplitudes on the order of 0.005--0.008 mag, are far smaller than the modulation amplitudes seen in the Blazhko effect \citep{Blazhko1907,Shapley1916}, yet their cumulative influence can produce quasi-periodic or irregular features in O-C diagrams \citep{Benko2025}. 

Interestingly, panels (c) and (d) in Figs.~\ref{fig:spec_res} and \ref{fig:spec_res_add} reveal the origin of the amplitude and frequency variations encoded in the light curves: they are concentrated mainly around the onset and in the tail of the hump structure.
Given that C2C variations are random and widespread in non-Blazhko RR Lyrae stars \citep{Benko2019,Benko2025}, the significant amplitude and frequency variations in the harmonics at the tail end of the hump structure (all samples in Figs.~\ref{fig:spec_res} and \ref{fig:spec_res_add}) can be considered as their origin, produced by fluctuating nonlinear responses of the turbulent convection to the pulsating wave.
This is reminiscent of the SX Phe star XX Cyg, where harmonic variations become increasingly erratic at the highest orders \citep{Niu2023}. 
This behavior may be universal in pulsating stars with convective envelopes, although further confirmation is required.

Based on the framework outlined in the previous subsection, the significant amplitude and frequency variations of the harmonics around the onset of the hump structure (see Fig.~\ref{fig:spec_res}) could be ascribed to the convection-pulsation (CP) interaction at the base of the convective envelope.
If the CP interaction operates on a longer timescale, lower-order harmonics can exhibit significant amplitude and frequency variations, which might correspond to the Blazhko effect, the long-standing mystery of quasi-periodic modulation in RR Lyrae stars.
This raises the possibility that the question posed by \citet{Kovacs2018}---``Are all RR Lyrae stars modulated?''—might receive an affirmative answer in principle, albeit with very small modulation amplitudes in non-Blazhko stars. However, this interpretation remains tentative. The modulation is not an all-or-nothing phenomenon; it could be universal, but its amplitude varies, and in non-Blazhko RR Lyrae stars, it may be only revealed by the exquisite sensitivity of their own harmonics, whose modulation is too weak to be detected by conventional means. A definitive conclusion will require dedicated observations of a larger sample with higher signal-to-noise ratios.

If this is the case, the harmonic variations around the onset of the hump structure (Blazhko effect, dominated by time-dependent CP interaction) and those in the tail (C2C variations, dominated by small-scale perturbations of the turbulent convection) are connected. Both are determined by the strength of turbulent convection in the envelope, a connection supported by the strong positive correlation between the C2C variation strength and the amplitude of the frequency-modulation component of the Blazhko effect in Blazhko RR Lyrae stars \citep{Benko2026}.

\subsection{Disharmonized Harmonics}
\label{sec:disharmonized}

In the conventional view, harmonics are completely determined by their parent pulsation mode, which would imply stable amplitudes and frequencies for the harmonics relative to those of the parent mode. 
However, this view is challenged not only by disharmonized harmonics in high-amplitude $\delta$ Scuti stars, but also by the harmonic detuning effects (HDE) observed in some Blazhko RR Lyrae stars \citep{Benko2018a}, which refer to a shift of a harmonic's observed frequency from the exact integer multiple ($nf_0$) of the parent frequency $f_0$. 

Theoretically, disharmonized harmonics would inevitably lead to HDE if the uncorrelated frequency variations are sufficiently large.
For simplicity, we generalize the concept of disharmonized harmonics from those that show uncorrelated amplitude, frequency, or phase variations (compared to their parent mode) to any harmonic whose amplitude, frequency, or phase (including their values and temporal variations) are not fully determined by its parent mode. 

According to the conventional view, $\Delta A/\bar{A}$ should remain constant for harmonics of different orders, and $\Delta f$ should increase linearly with the harmonic order. However, the observational results (see panels (c) and (d) in Figs.~\ref{fig:spec_res} and \ref{fig:spec_res_add}) clearly show otherwise, suggesting that (generalized) disharmonized harmonics are universal in non-Blazhko RRab stars and that additional nonlinear processes are involved in their generation.

\section{Conclusions}
\label{sec:conc}

In this work, we have explored the hump structures formed by the harmonics of the primary pulsation modes in the frequency spectra of non-Blazhko RRab stars, using {\it Kepler} SC photometric data.
An interesting relationship has emerged: the harmonics near the onset of the hump structure exhibit significant amplitude and frequency variations in a substantial fraction of the sample, while those in the tail of the hump structure show such variations in all cases.

We have proposed a framework in which the hump structure may be associated with the turbulent convective envelope of the star. Within this framework (which remains a working hypothesis), several important phenomena observed in these pulsating stars (including HDE, C2C variations, disharmonized harmonics, and the hump structure itself) can be explained self-consistently. While the proposed interpretation requires validation through hydrodynamic modeling, the identified harmonic hump and its associated variability constitute a robust observational finding that any successful model of convection-pulsation interaction must explain.

Our findings demonstrate that harmonics can act as a microscope on stellar variability, serving as in-situ probes of convective dynamics rather than passive Fourier components, and thus provide a powerful new diagnostic for one of stellar astrophysics' most enduring challenges.

\begin{acknowledgements}
We would like to thank the anonymous referee for professional and significant suggestions for improving the work. J.S.N. is grateful to Jue-Ran Niu for a supportive and productive working environment.
The author acknowledges the \textit{Kepler} Science Team and all individuals who contributed to the success of the \textit{Kepler} mission.
All \textit{Kepler} data used in this paper are publicly available through the Mikulski Archive for Space Telescopes (MAST).
\end{acknowledgements}


\clearpage
\setcounter{figure}{0}
\setcounter{table}{0}
\renewcommand{\thefigure}{A\arabic{figure}}
\renewcommand{\thetable}{A\arabic{table}}
\onecolumn
\begin{appendix}

\section{Additional Results}
\label{app:add_results}

\begin{figure*}[htbp!]
  \centering
  \includegraphics[width=0.46\textwidth]{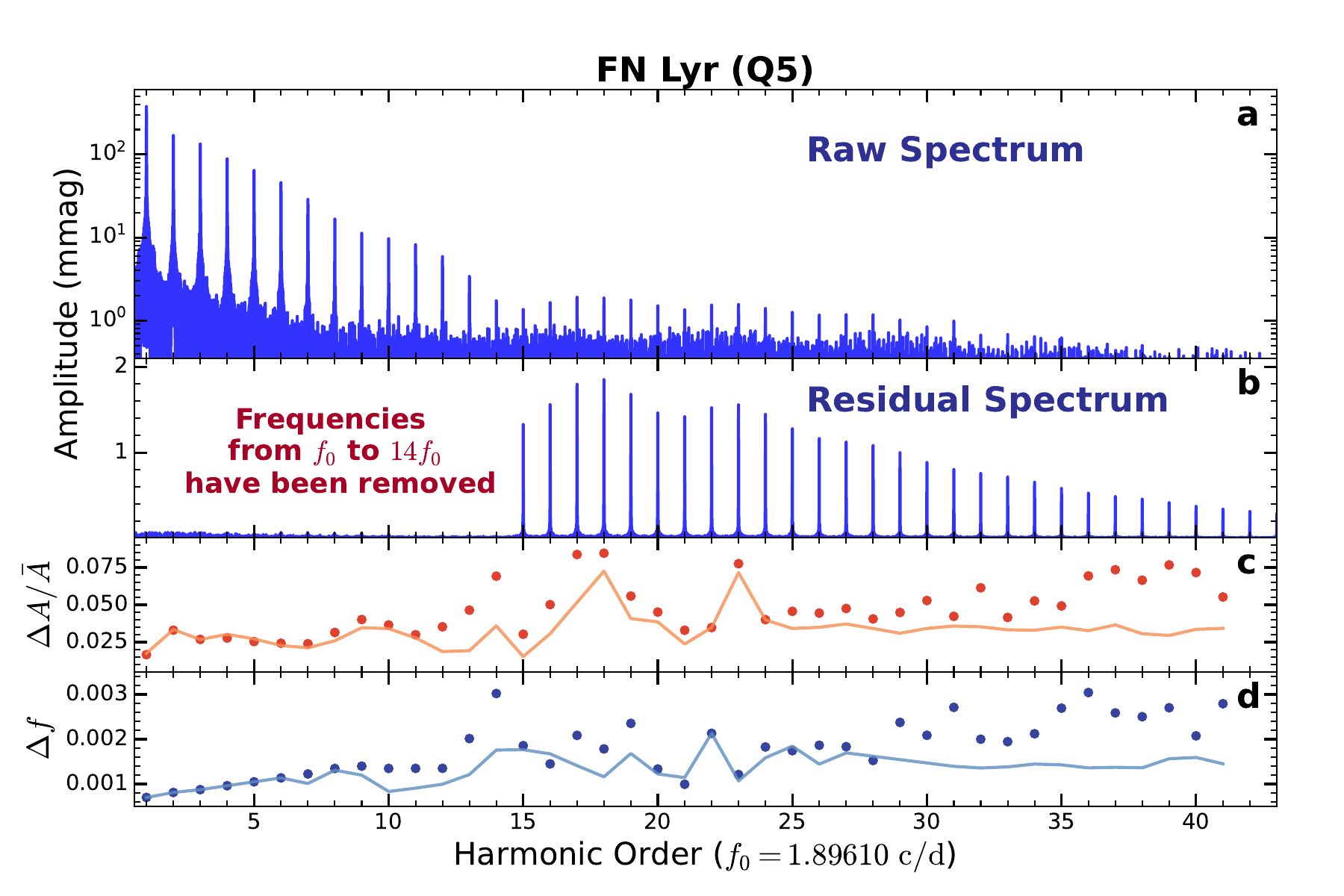}
  \includegraphics[width=0.46\textwidth]{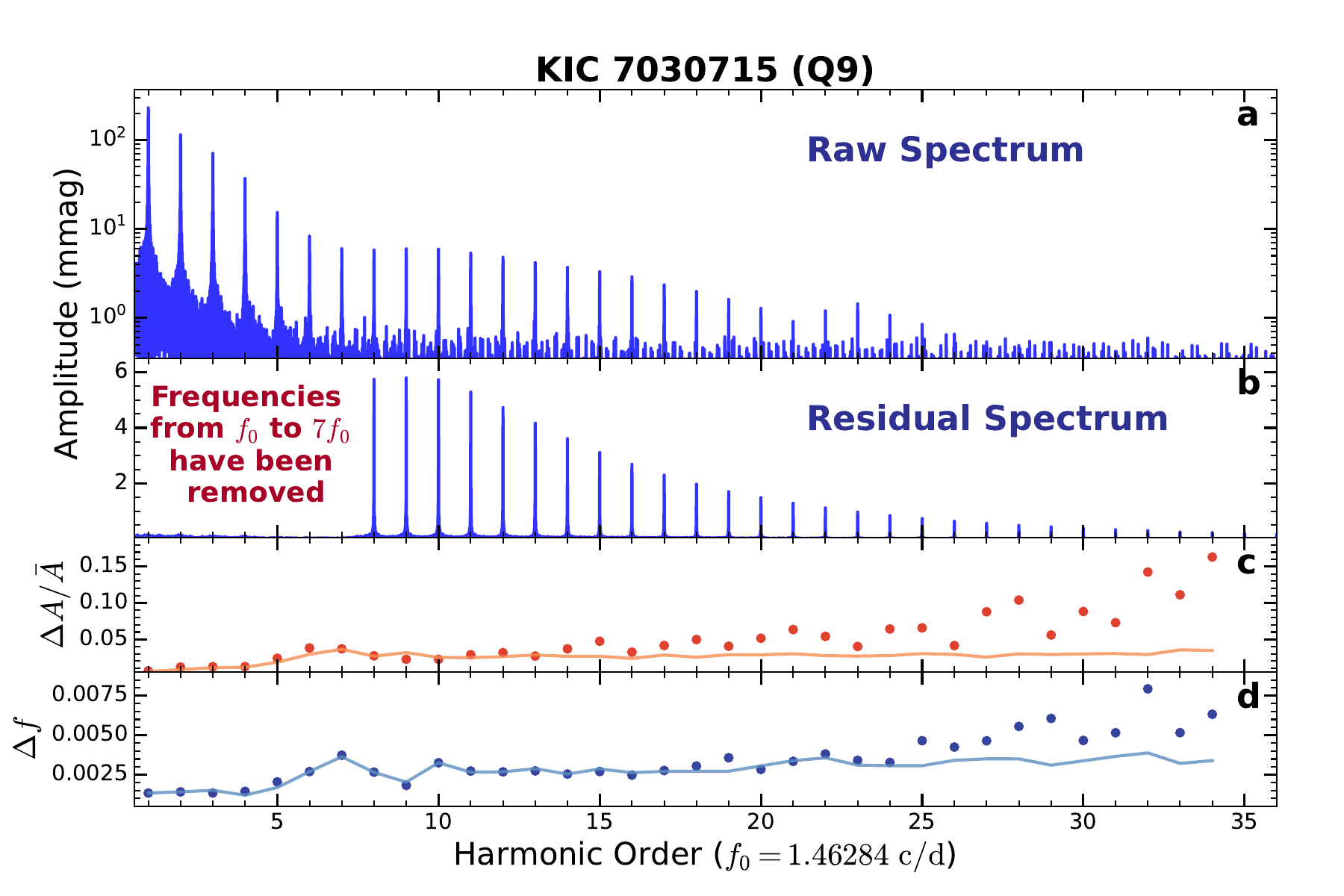}
  \includegraphics[width=0.46\textwidth]{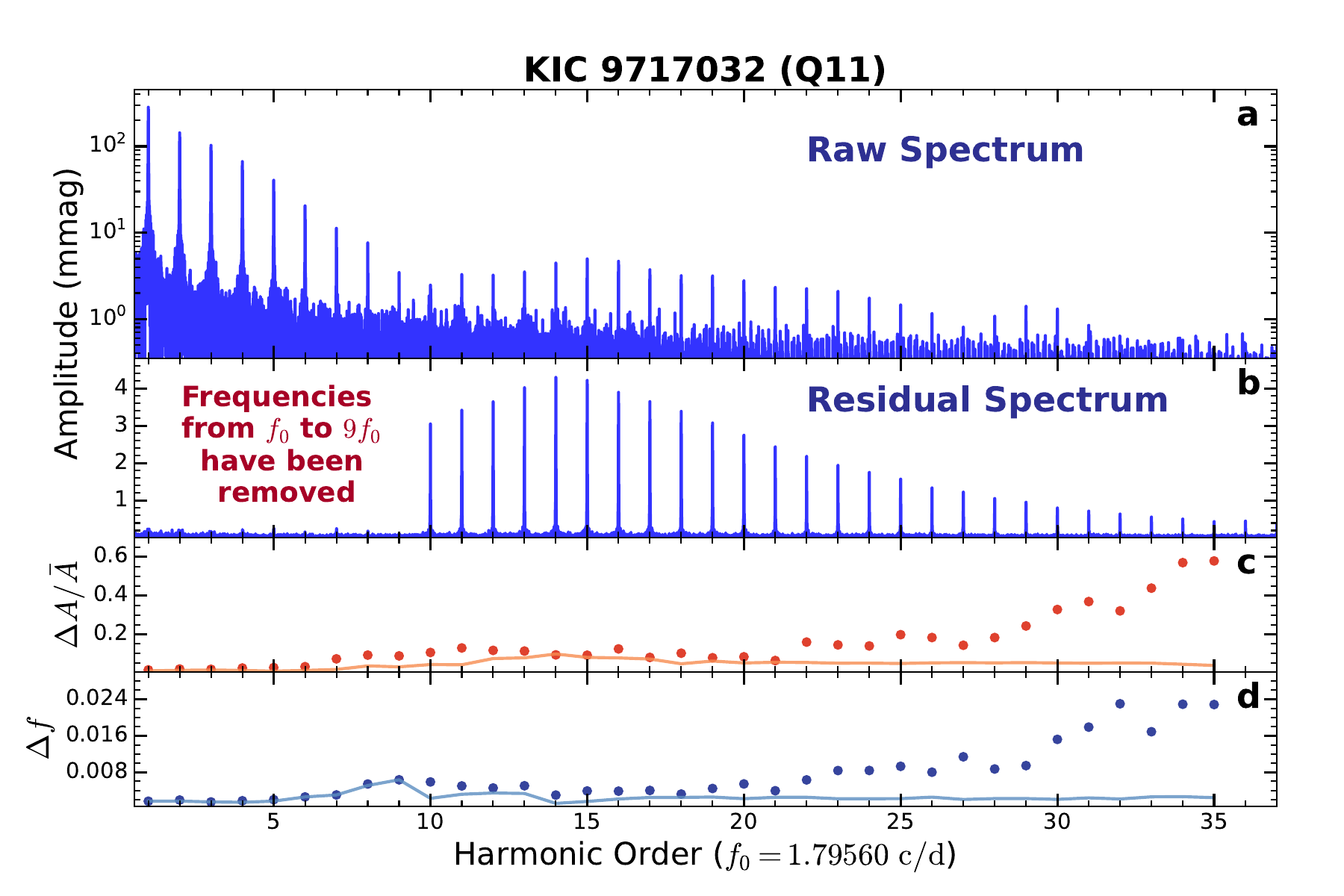}
  \includegraphics[width=0.46\textwidth]{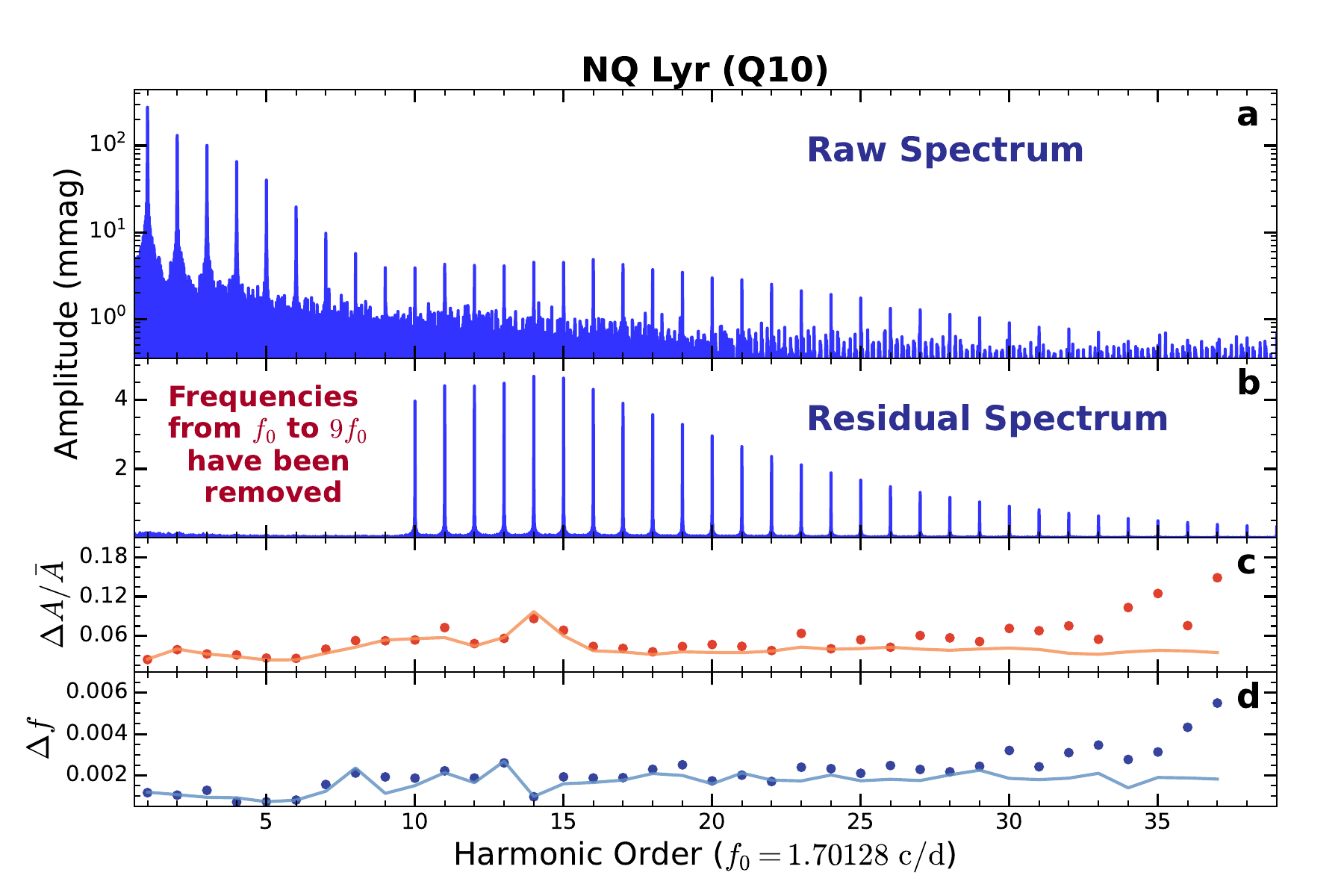}
  \includegraphics[width=0.46\textwidth]{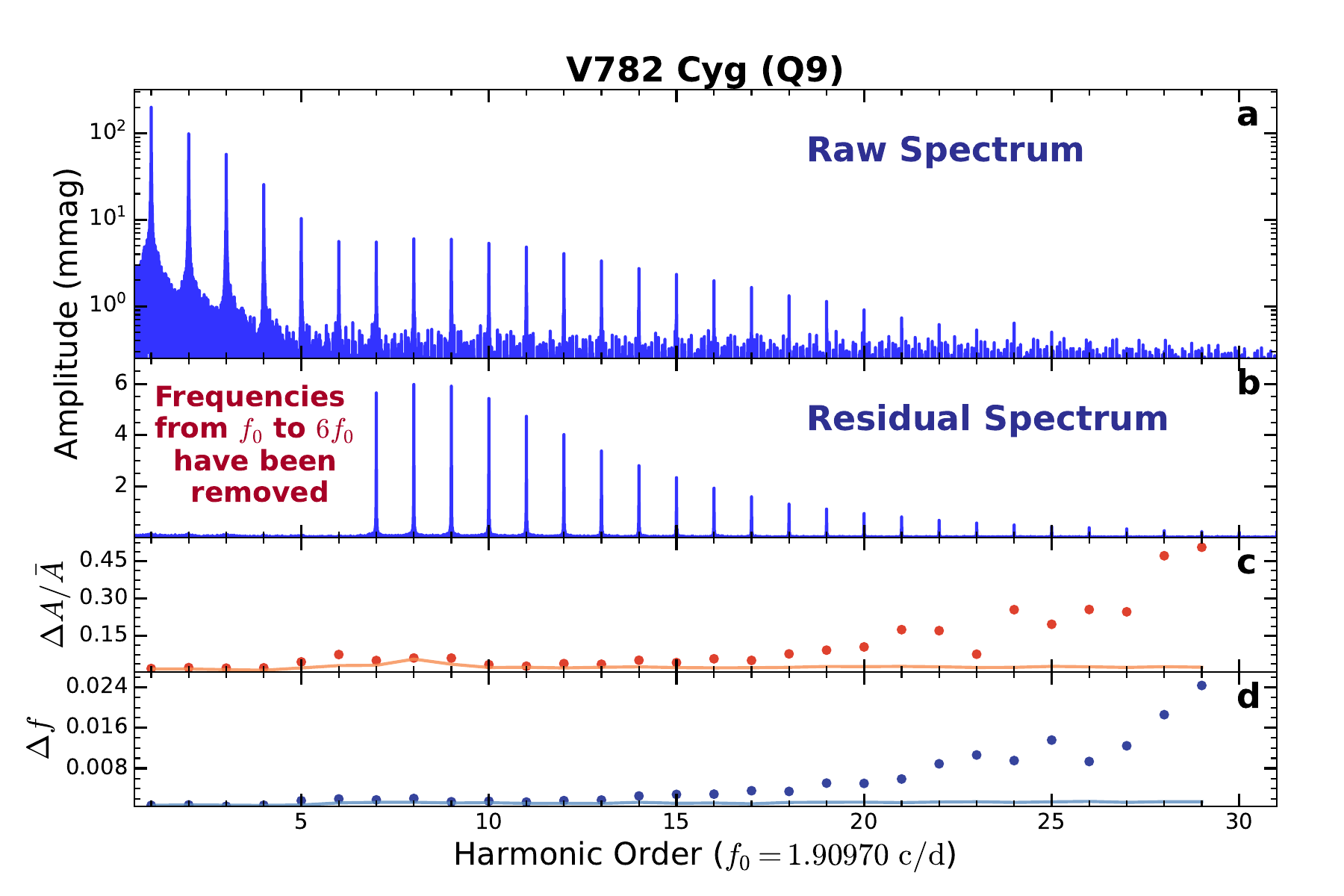}
  \includegraphics[width=0.46\textwidth]{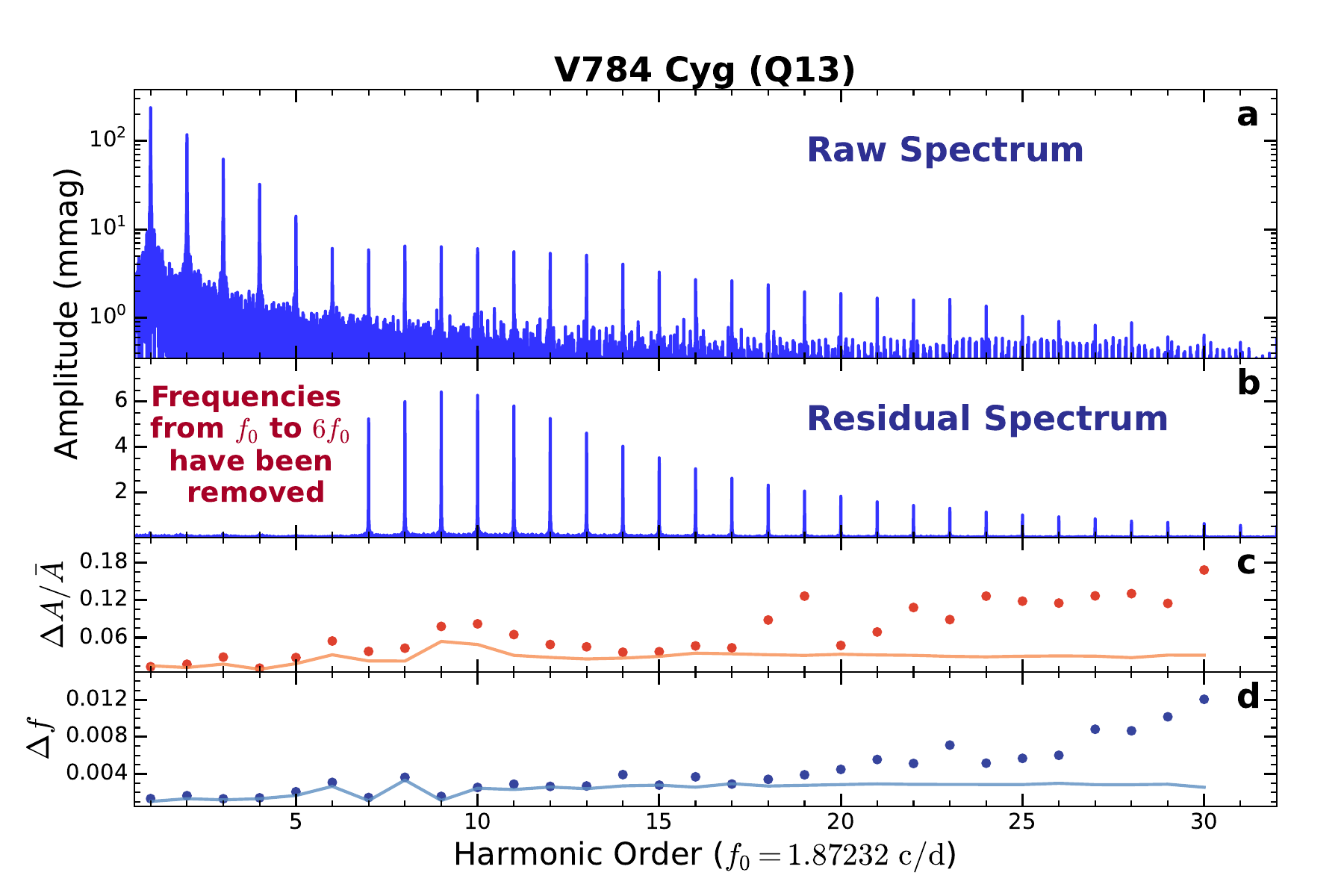}
  \includegraphics[width=0.46\textwidth]{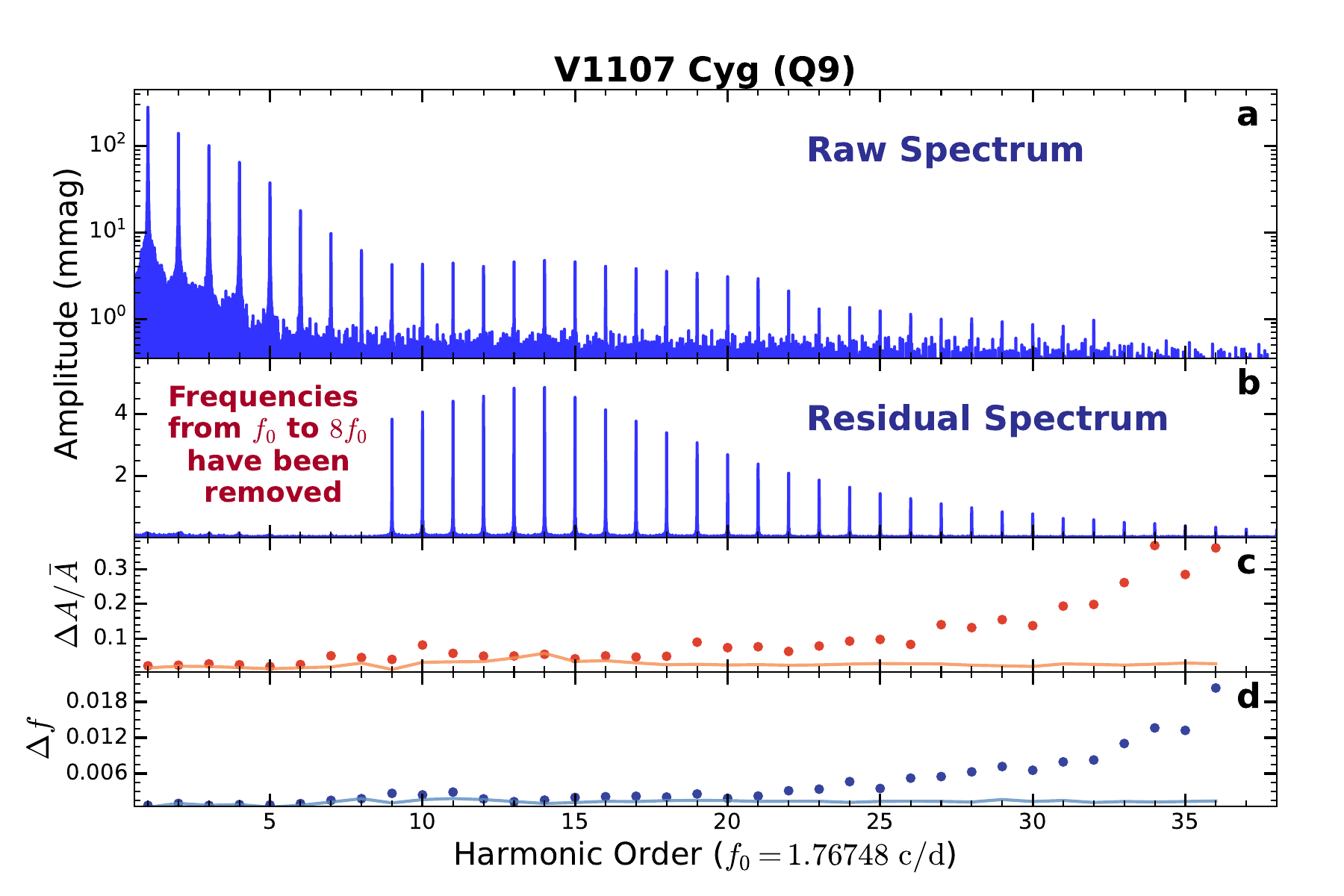}
  \includegraphics[width=0.46\textwidth]{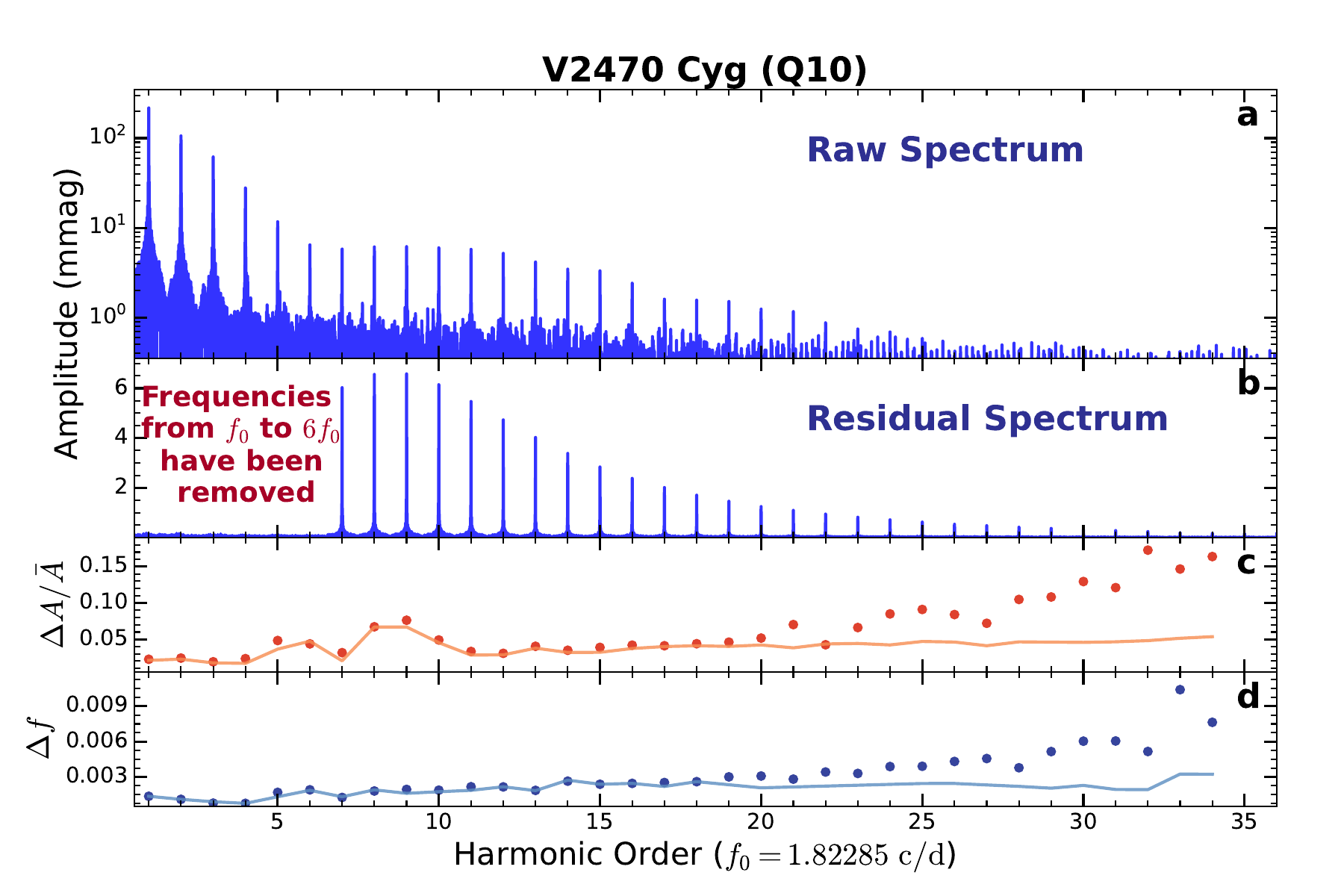}
  \caption{The harmonics in the frequency spectra and the magnitude of their amplitude and frequency variations (additional samples).}
  \label{fig:spec_res_add}
\end{figure*}

\clearpage
\section{Data Reduction}
\label{app:data_reduc}

This section provides a detailed account of the data reduction, harmonic extraction, and time-series analysis procedures that underpin the main findings reported in this work.

\subsection{Extraction of Harmonic Amplitudes and Frequencies}
\label{app:methods01}

To extract the amplitudes and frequencies of the harmonic pulsation modes, we performed a standard pre-whitening procedure. At each step, the Lomb-Scargle algorithm \citep{Lomb_Scargle} was used to identify the frequency of the highest-amplitude signal. This was followed by a nonlinear least-squares fit to refine the frequency and determine the corresponding amplitude and phase. The best-fitting sinusoidal model was then subtracted from the light curve, and the process was repeated on the residual data.
We continued this iterative pre-whitening until the signal-to-noise ratio (S/N) was less than 8.0. 

The S/N values quoted in the above pre-whitening procedure are local signal-to-noise ratios computed from the residual spectrum at each successive pre-whitening step. They reflect the significance of the peak relative to the immediate background noise at that stage of the analysis, and are not directly comparable across different harmonics or to the S/N in the original, unwhitened spectrum.

The uncertainties in all pre-whitening processes throughout this work were estimated using the framework introduced by \citet{Zong2016_sdb}. In this framework, the amplitude uncertainties ($\sigma_A$) are defined as the median value of the amplitudes within a Lomb-Scargle spectral window of $2\ \cd$ centered on the frequency peak, while the frequency uncertainties ($\sigma_f$) are estimated following the formalism proposed by \citet{Montgomery1999} and \citet{Aerts2021}.

\subsection{Time-Resolved Analysis: Short-Time Fourier Transform}
\label{app:methods02}

A sliding window of 15 days in length was advanced across the entire dataset in increments of 1.5 days. For each window position, we repeated the full pre-whitening procedure described above, extracting the amplitudes and frequencies of all detectable harmonics (S/N > 8.0). The choice of a 15-day window represents a balance between temporal resolution and frequency precision, allowing us to resolve variations on timescales relevant to the phenomena of interest. The phase was treated as a free parameter in the fit but was not the primary focus of this analysis \citep{Xue2024}.

The STFT process was terminated when the S/N of a harmonic frequency fell below 8.0 in any 15-day window. From the extracted harmonics, we thus obtained a time series of amplitude and frequency values, with each measurement timestamped at the midpoint of its corresponding 15-day window. These time series form the basis for all subsequent analyses of harmonic variability.

\section{AW Dra as An Example}
\label{app:AWDra}

\begin{figure*}[htp]
  \centering
  \includegraphics[width=0.48\textwidth]{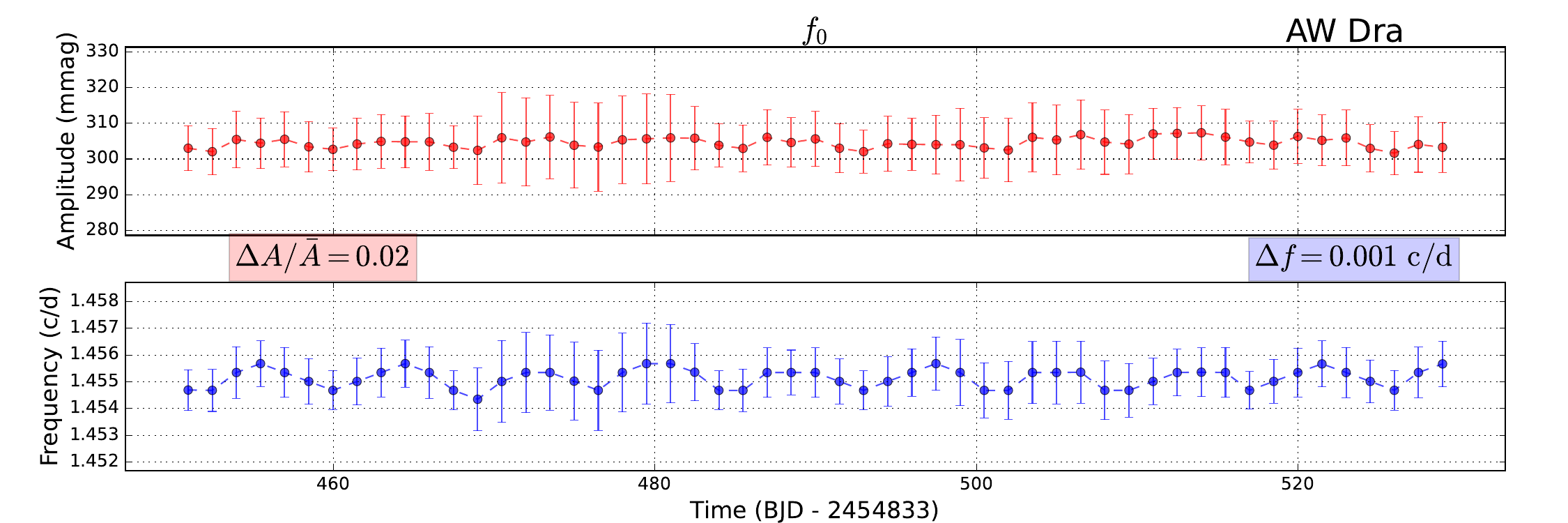}
  \includegraphics[width=0.48\textwidth]{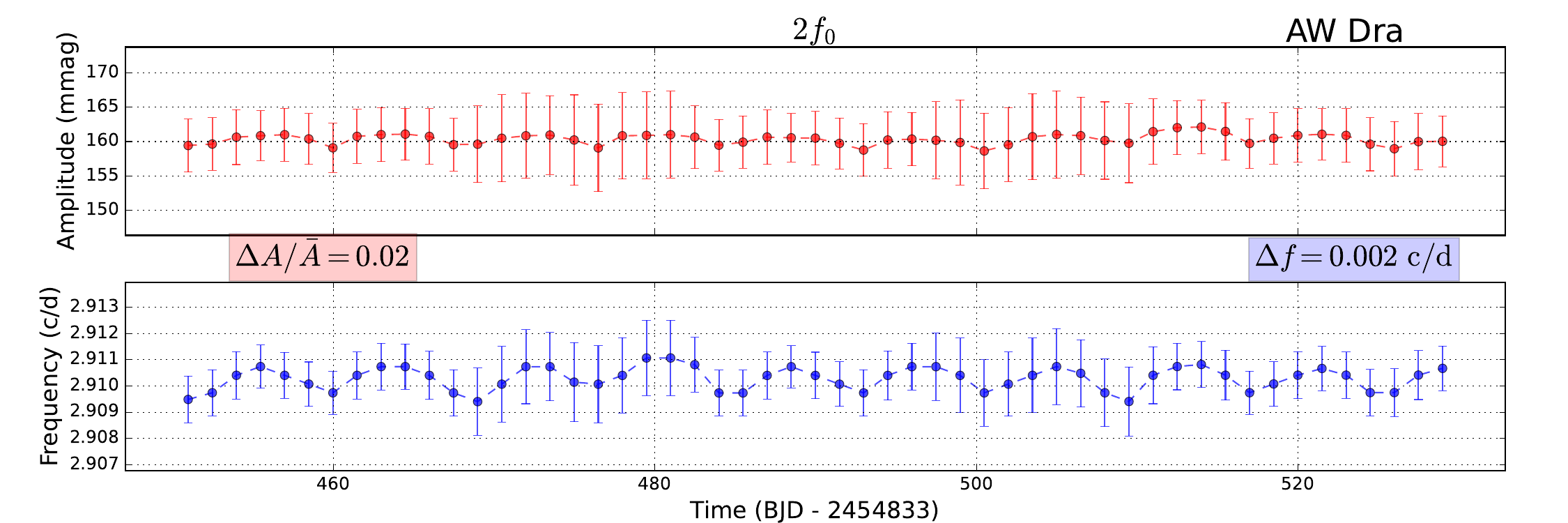}
  \includegraphics[width=0.48\textwidth]{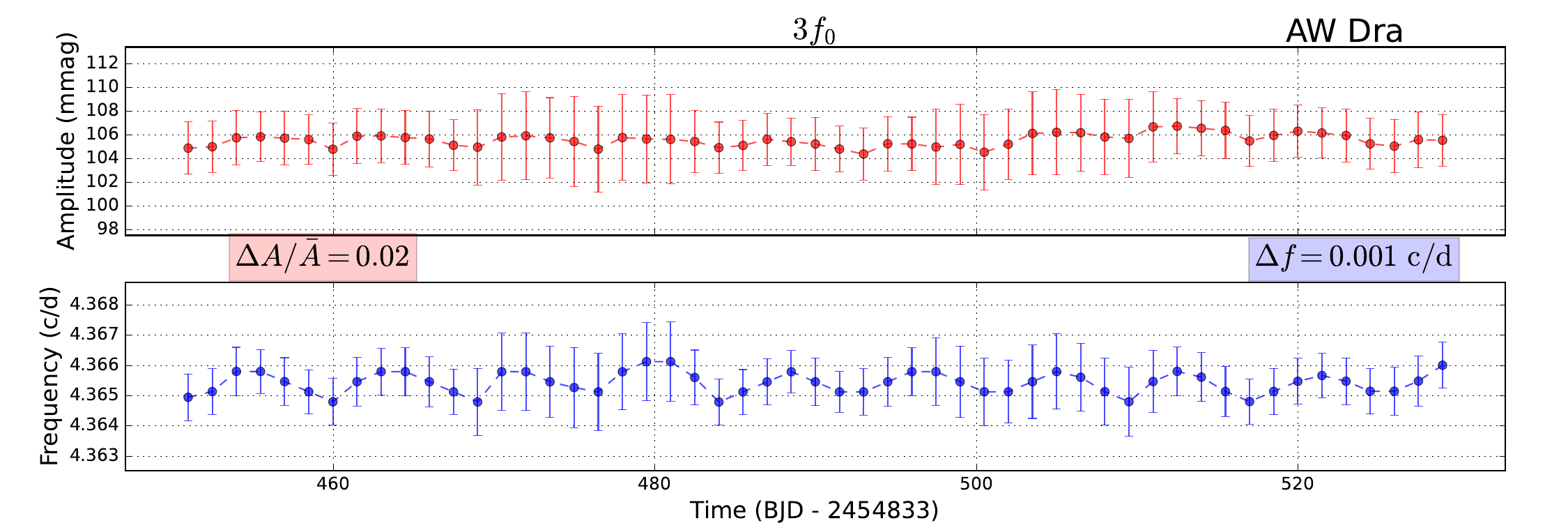}
  \includegraphics[width=0.48\textwidth]{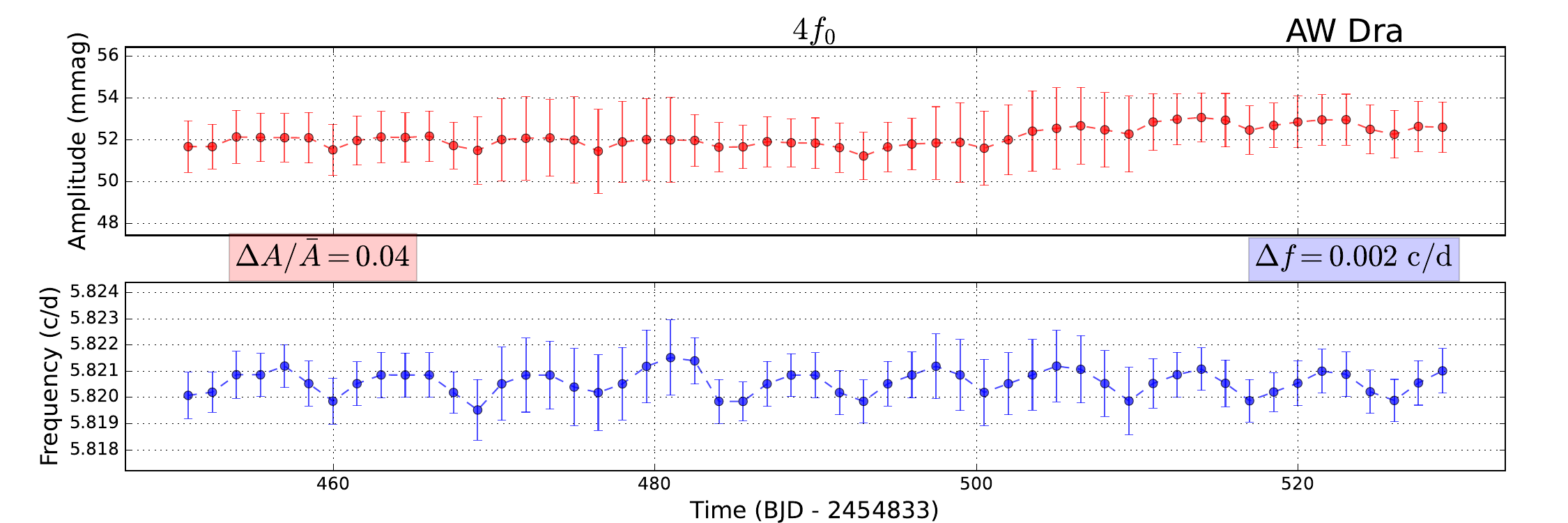}
  \includegraphics[width=0.48\textwidth]{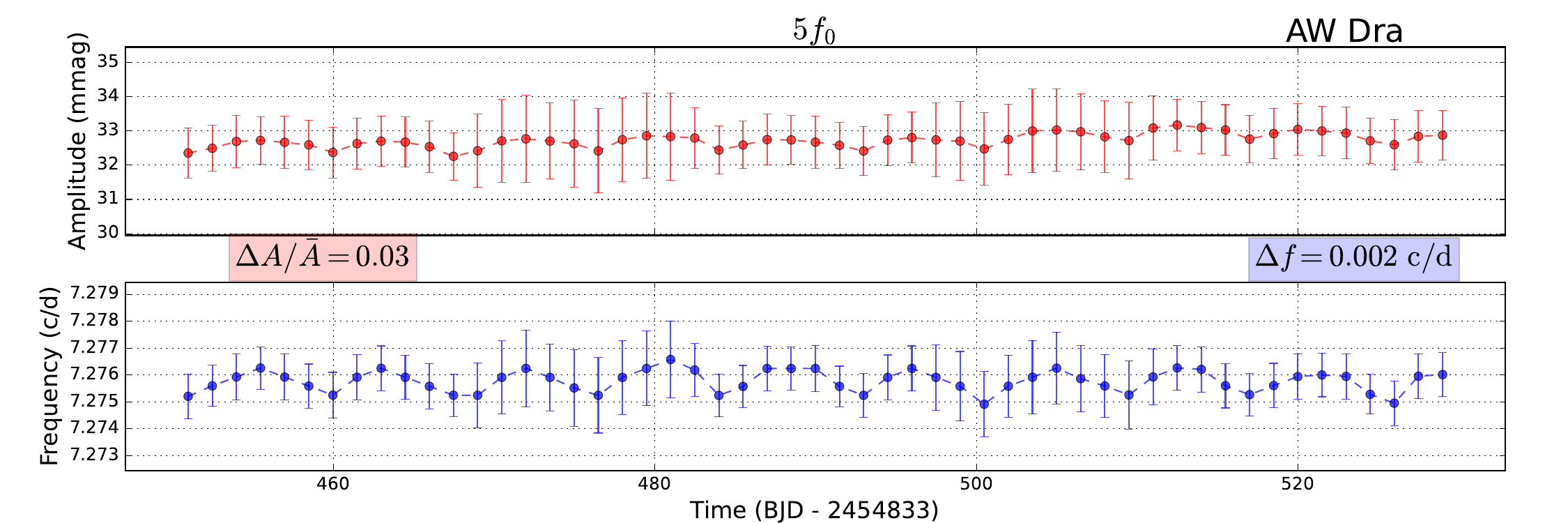}
  \includegraphics[width=0.48\textwidth]{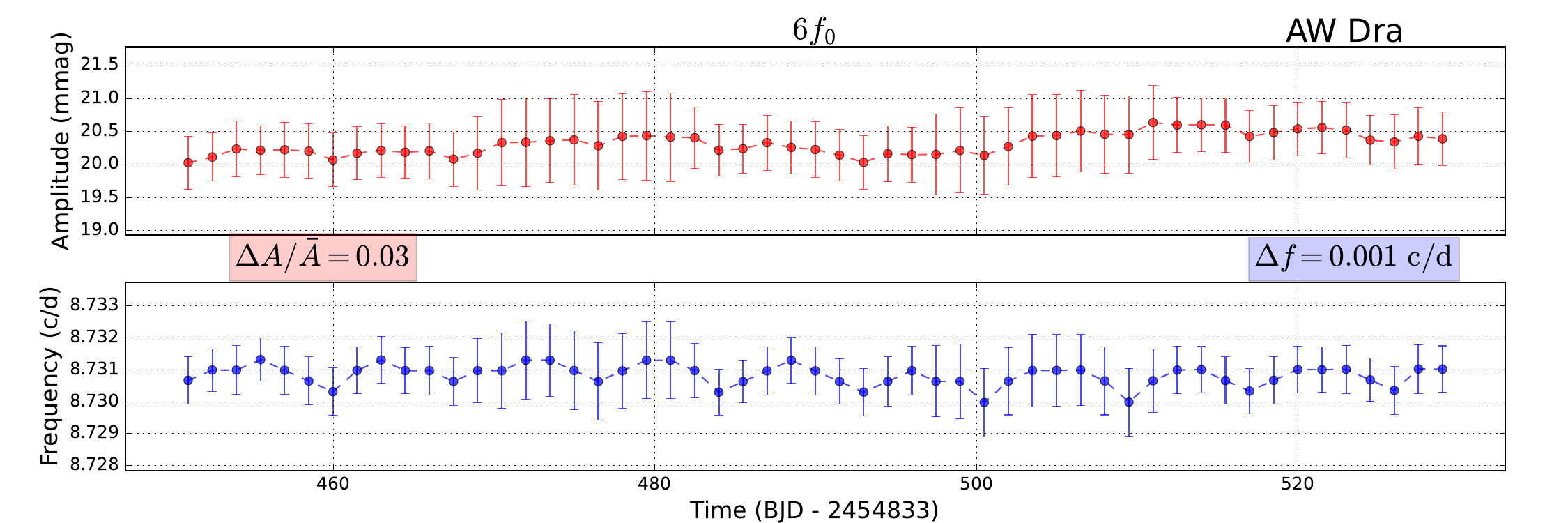}
  \caption{Temporal variations in amplitude and frequency for harmonics $f_0$--$36f_0$, part I.}
  \label{fig:var_amp_freq01}
\end{figure*}

\begin{figure*}[htp]
  \centering
  \includegraphics[width=0.48\textwidth]{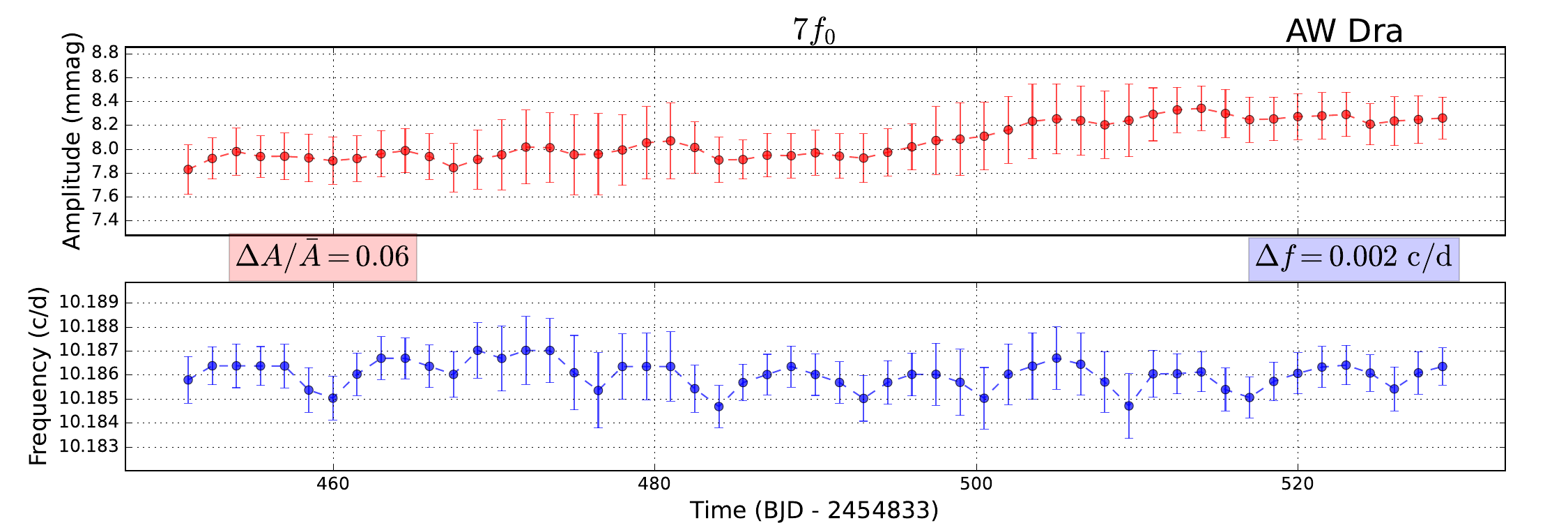}
  \includegraphics[width=0.48\textwidth]{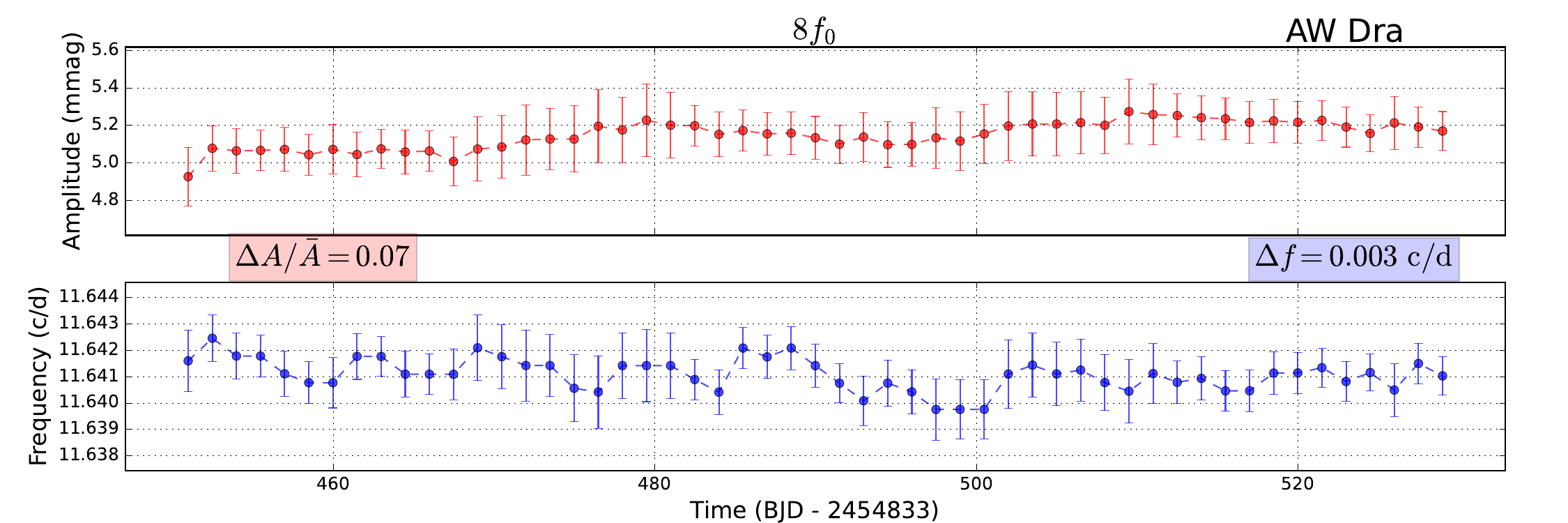}
  \includegraphics[width=0.48\textwidth]{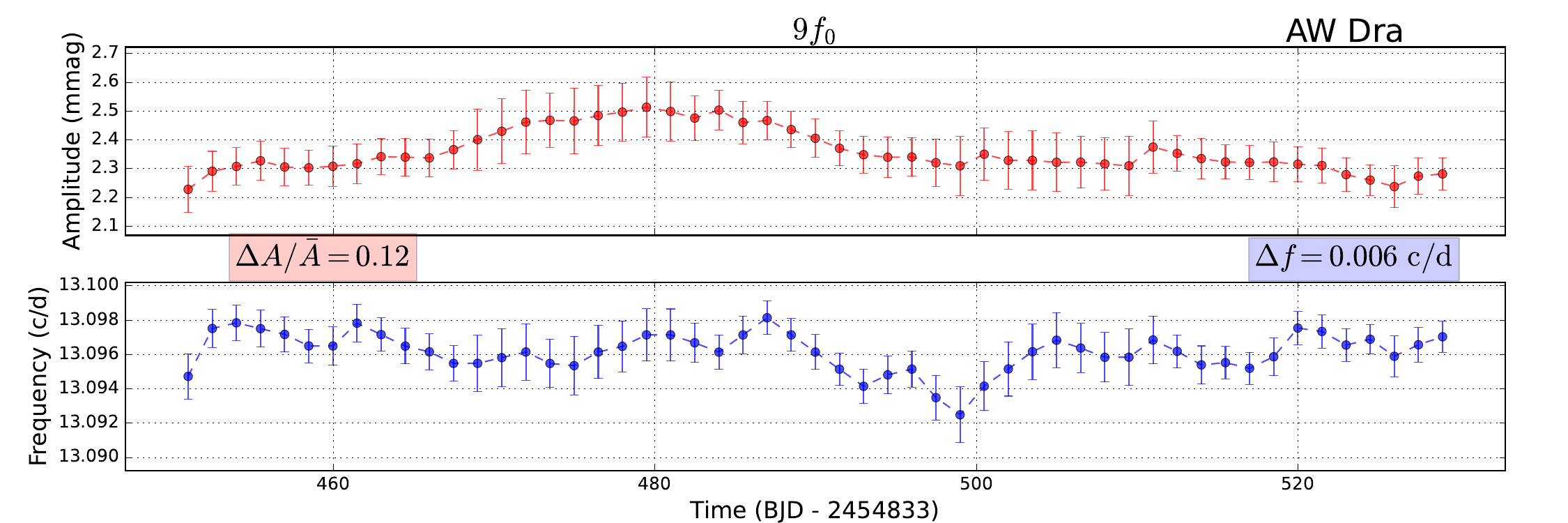}
  \includegraphics[width=0.48\textwidth]{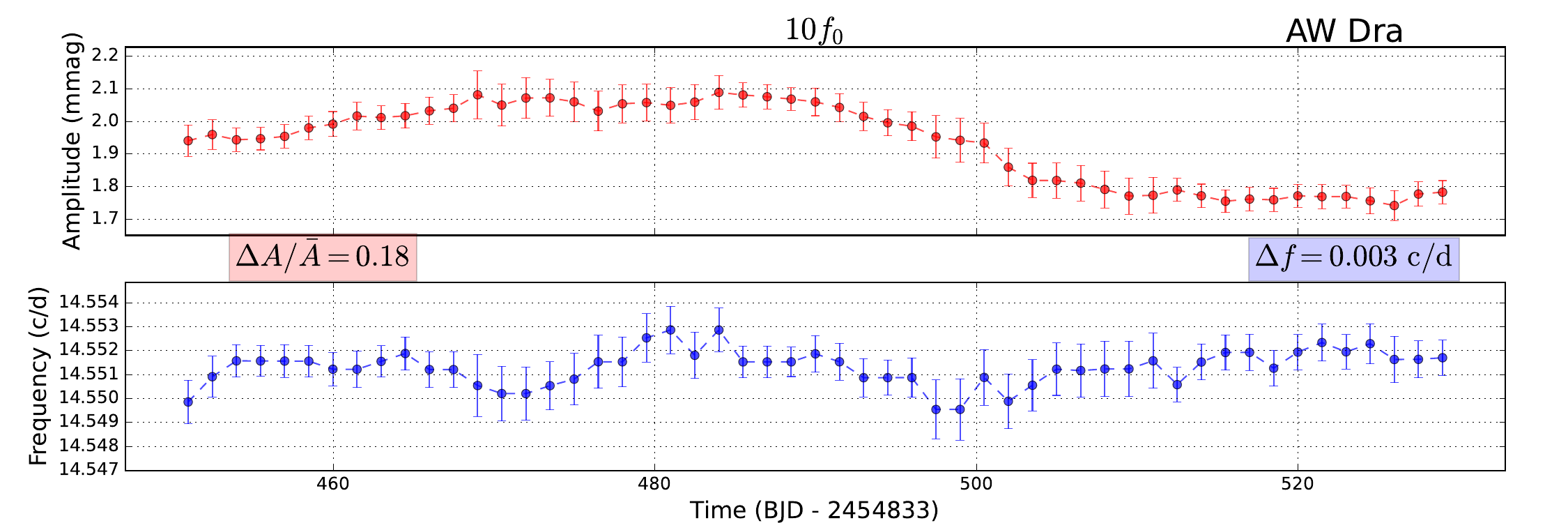}
  \includegraphics[width=0.48\textwidth]{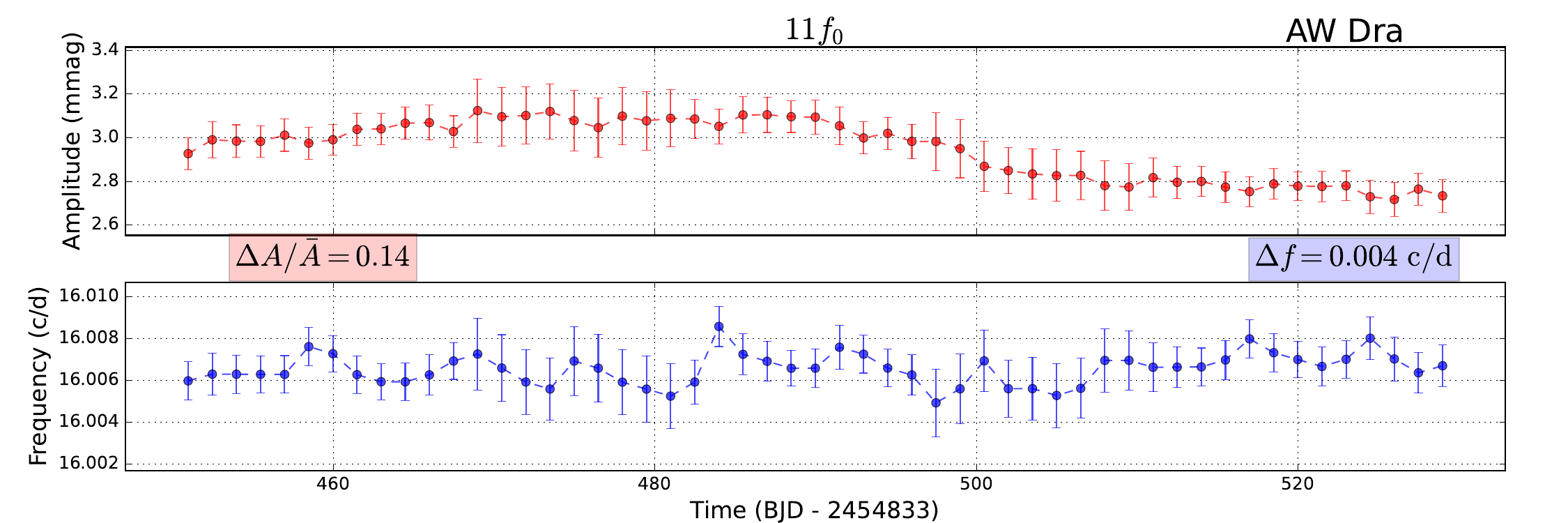}
  \includegraphics[width=0.48\textwidth]{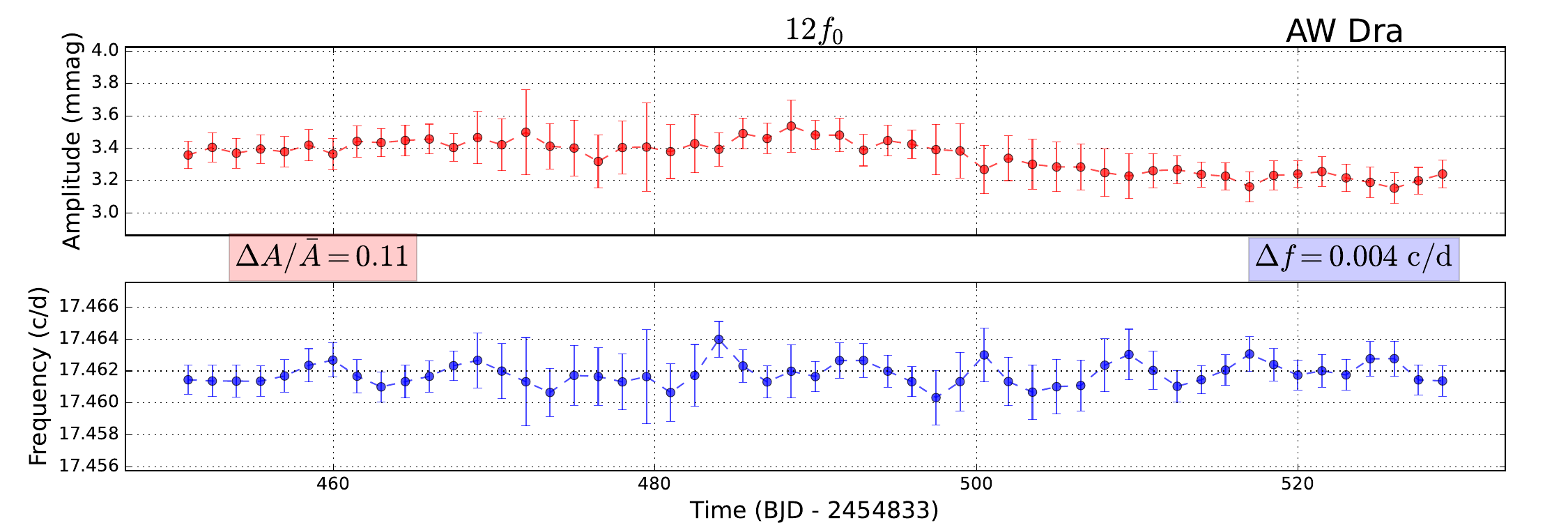}
  \includegraphics[width=0.48\textwidth]{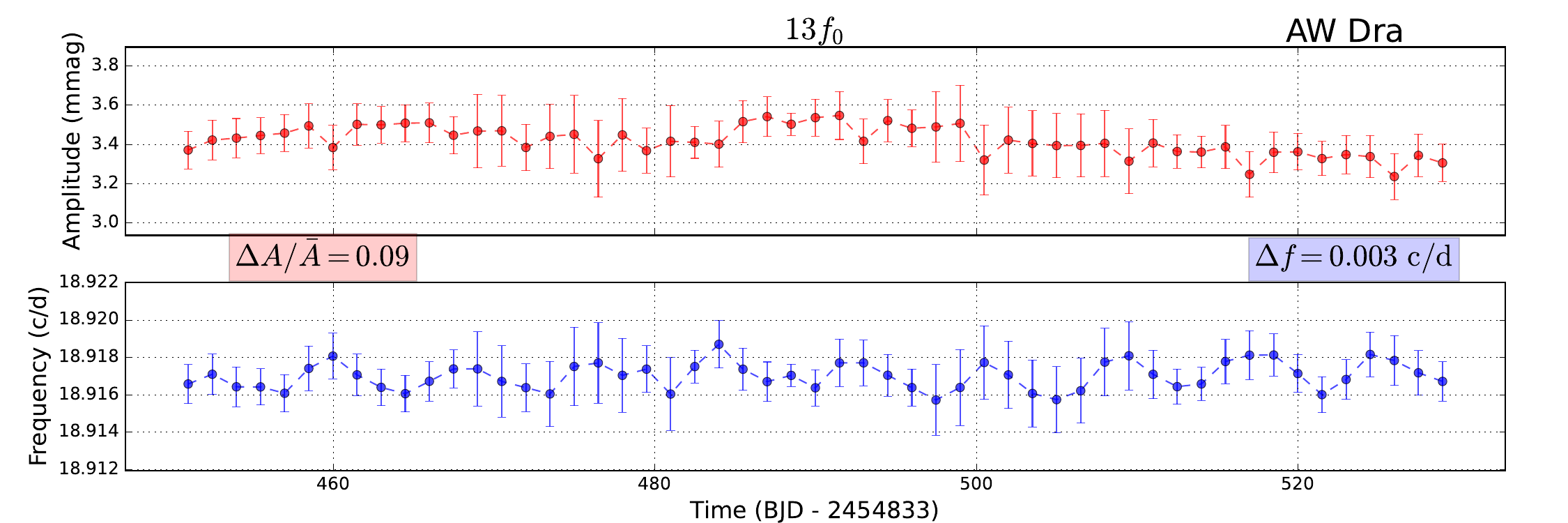}
  \includegraphics[width=0.48\textwidth]{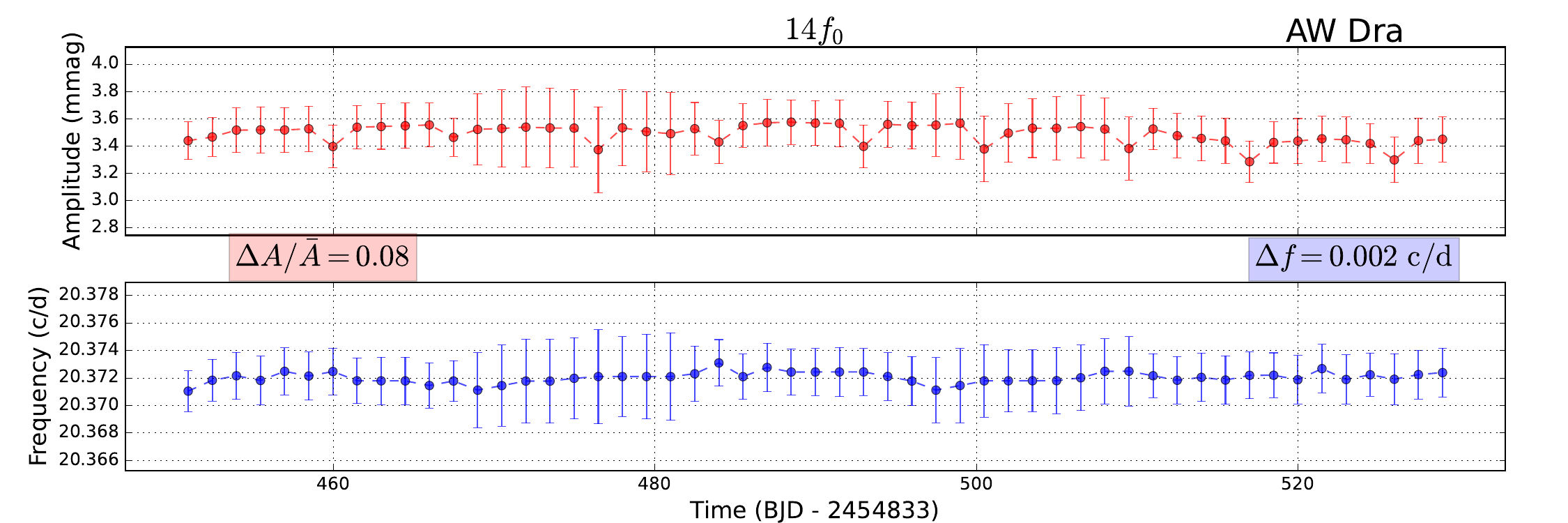}
  \includegraphics[width=0.48\textwidth]{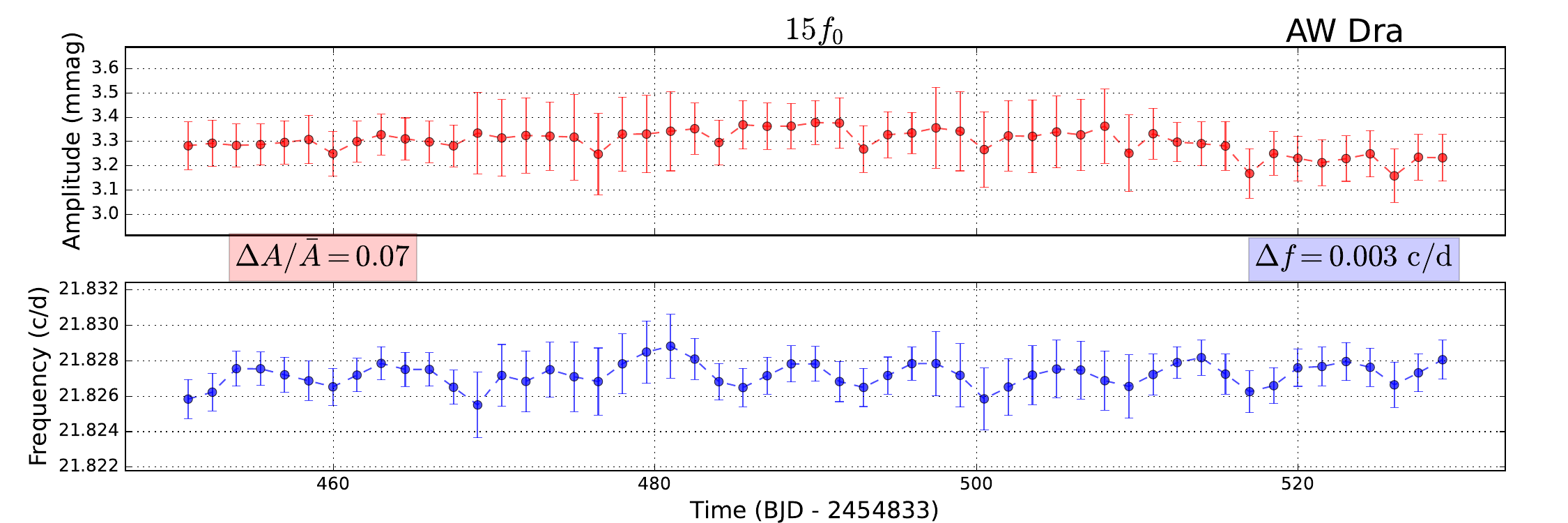}
  \includegraphics[width=0.48\textwidth]{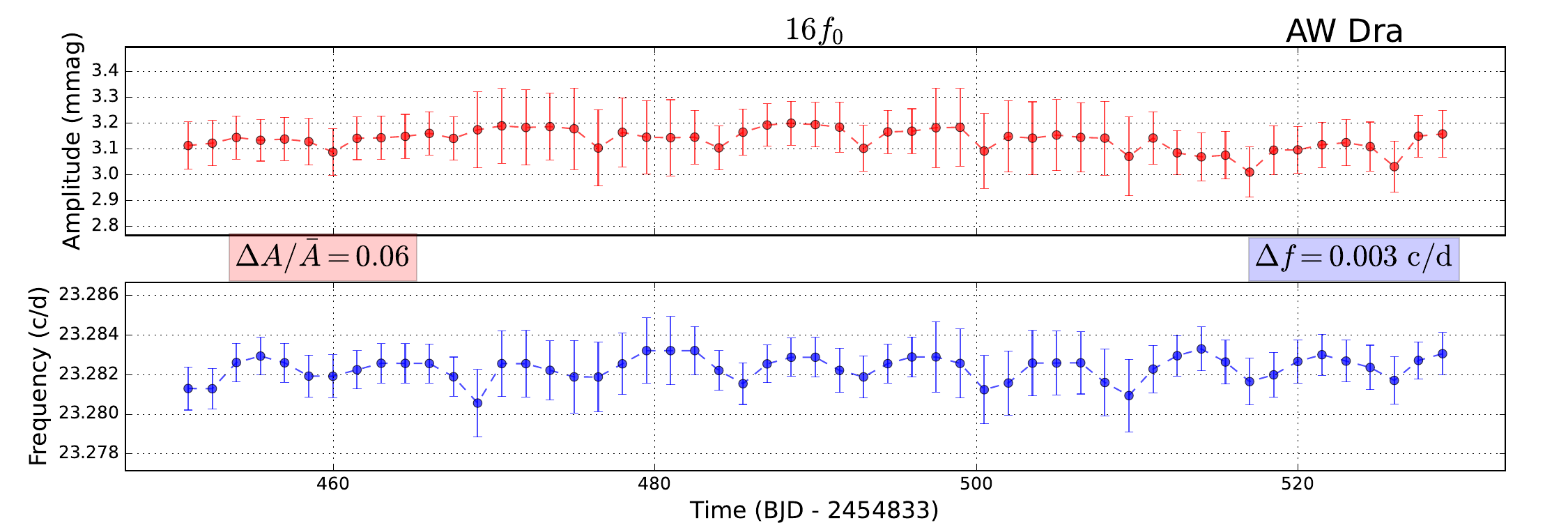}
  \includegraphics[width=0.48\textwidth]{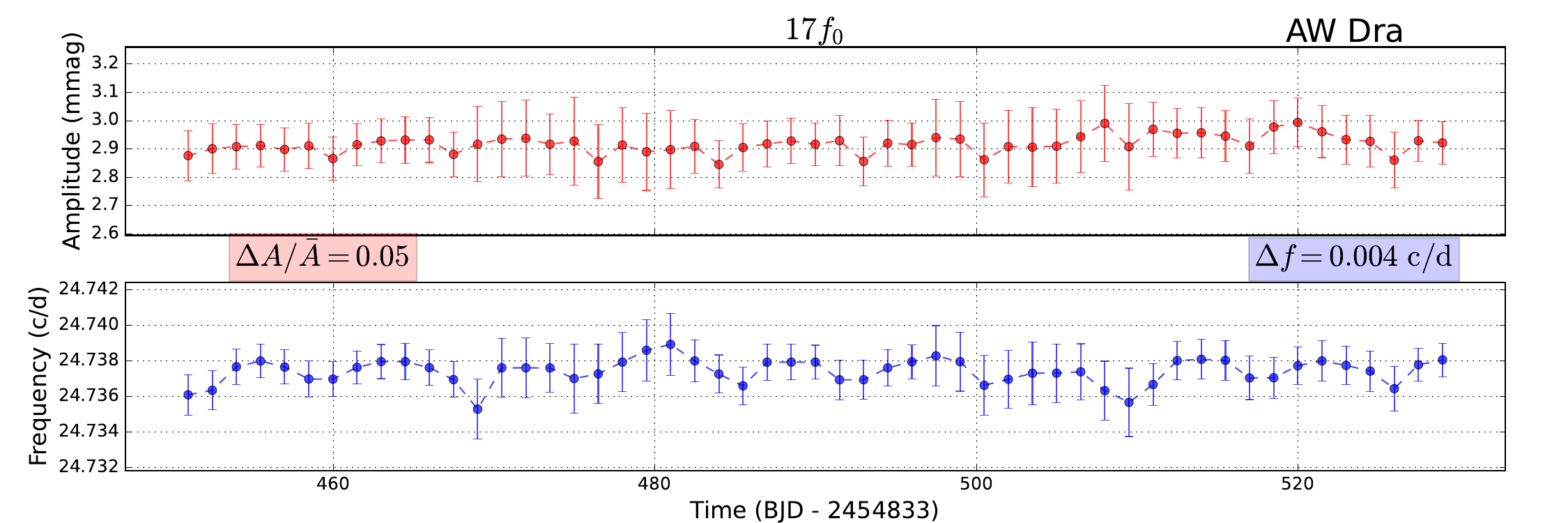}
  \includegraphics[width=0.48\textwidth]{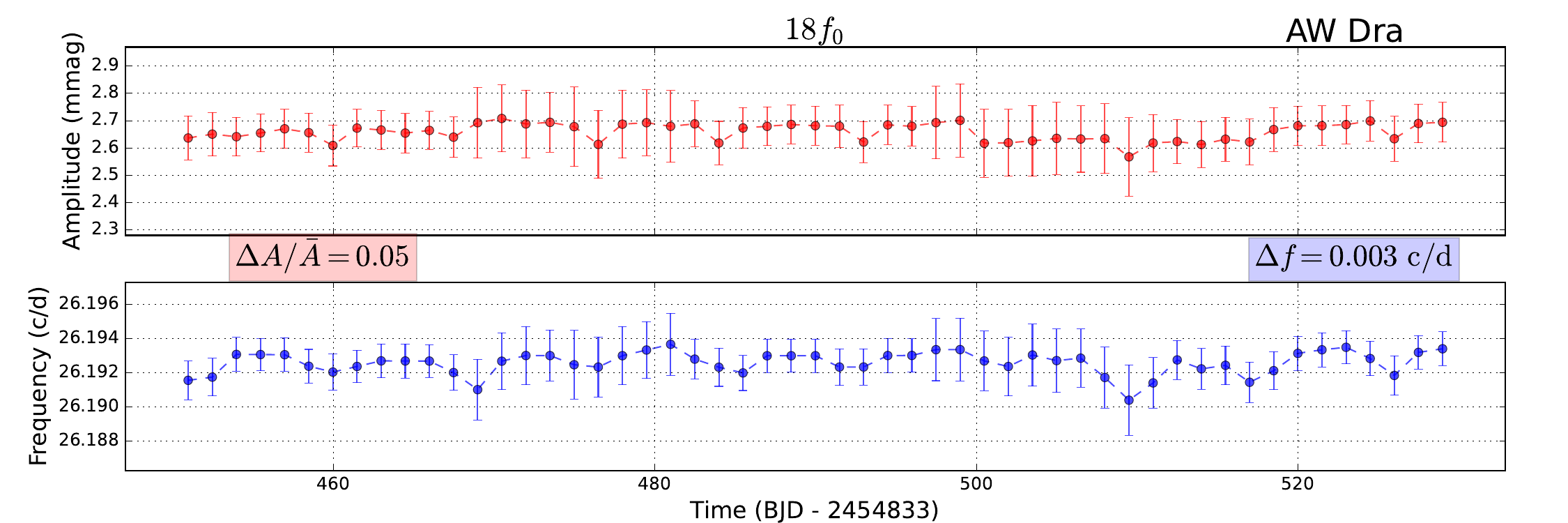}
  \includegraphics[width=0.48\textwidth]{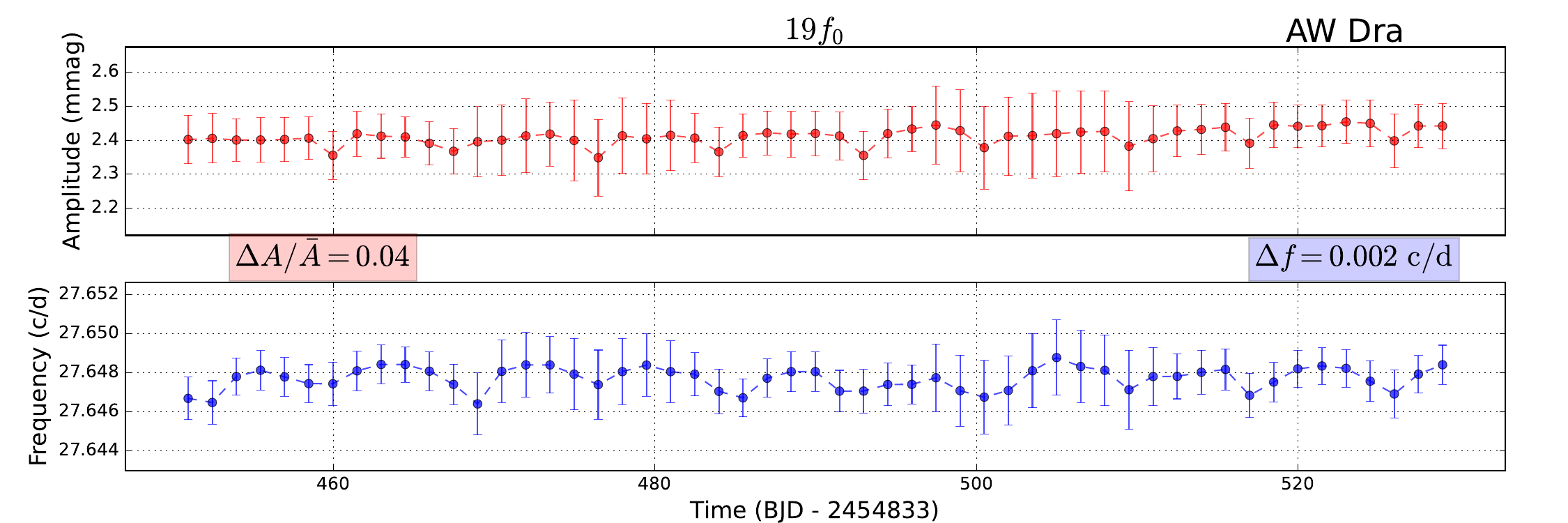}
  \includegraphics[width=0.48\textwidth]{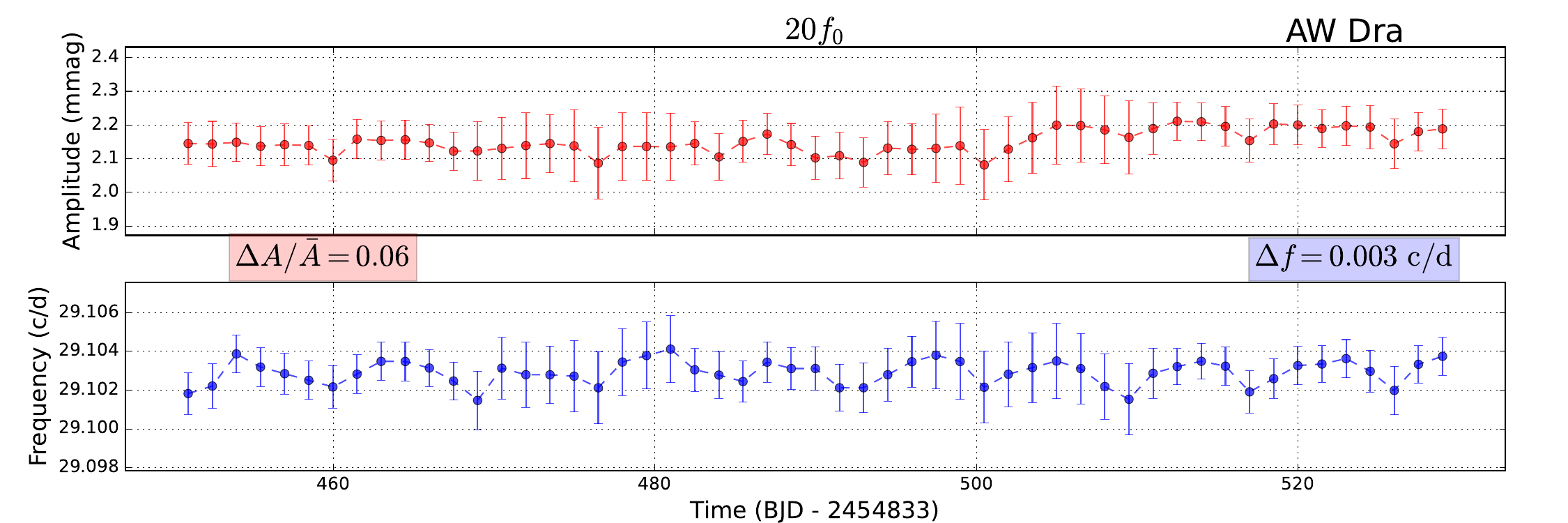}
  \includegraphics[width=0.48\textwidth]{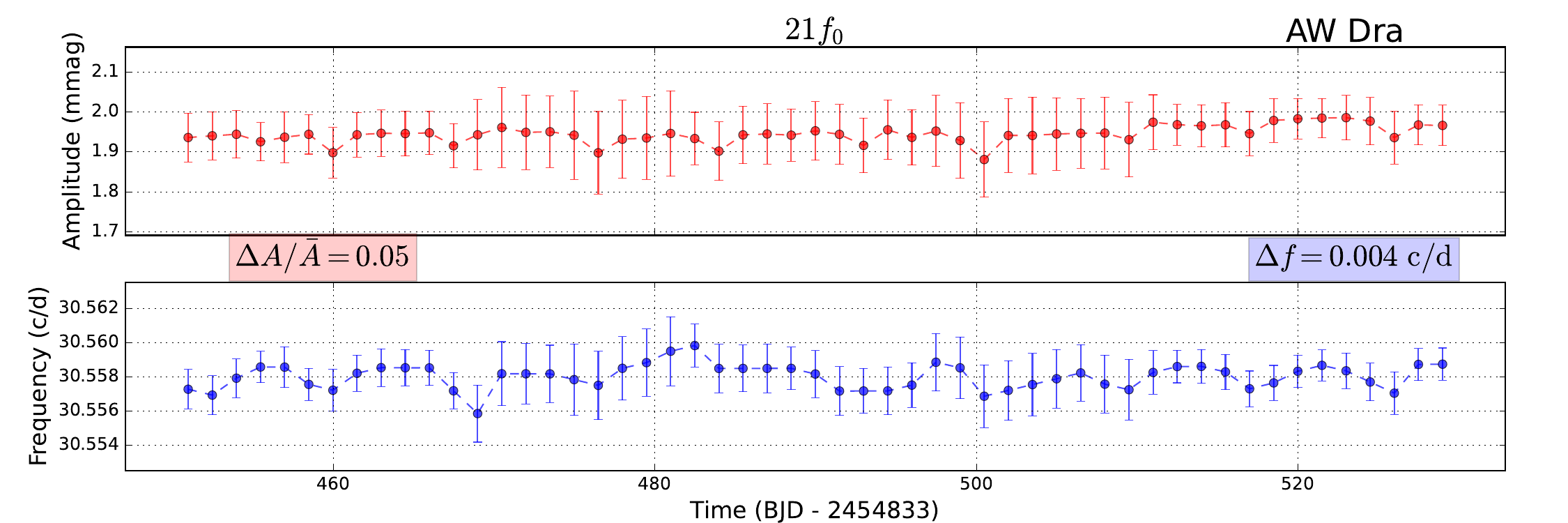}
  \includegraphics[width=0.48\textwidth]{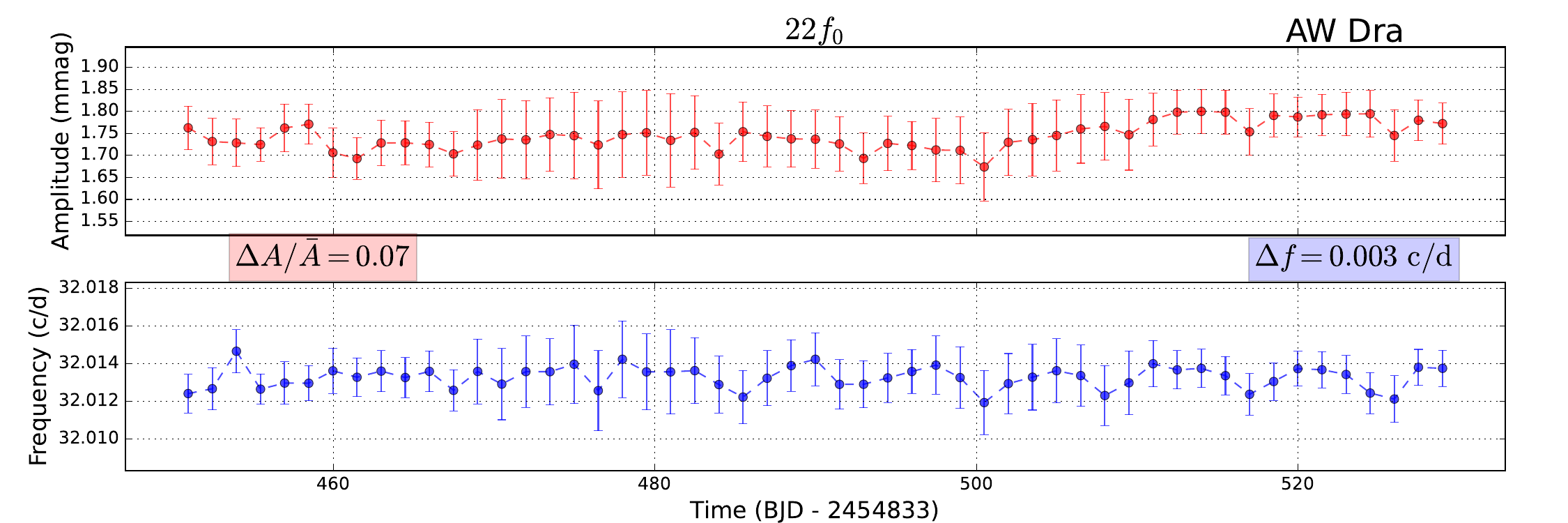}
  \caption{Temporal variations in amplitude and frequency for harmonics $f_0$--$36f_0$, part II.}
  \label{fig:var_amp_freq02}
\end{figure*}

\begin{figure*}[htp]
  \centering
  \includegraphics[width=0.48\textwidth]{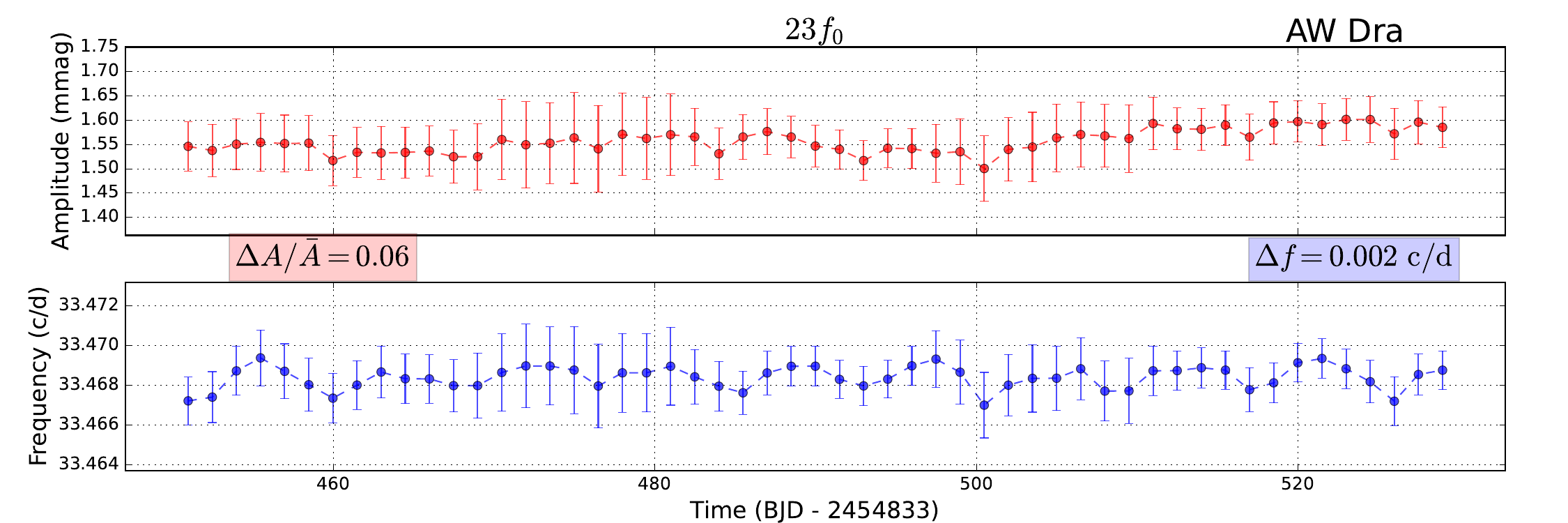}
  \includegraphics[width=0.48\textwidth]{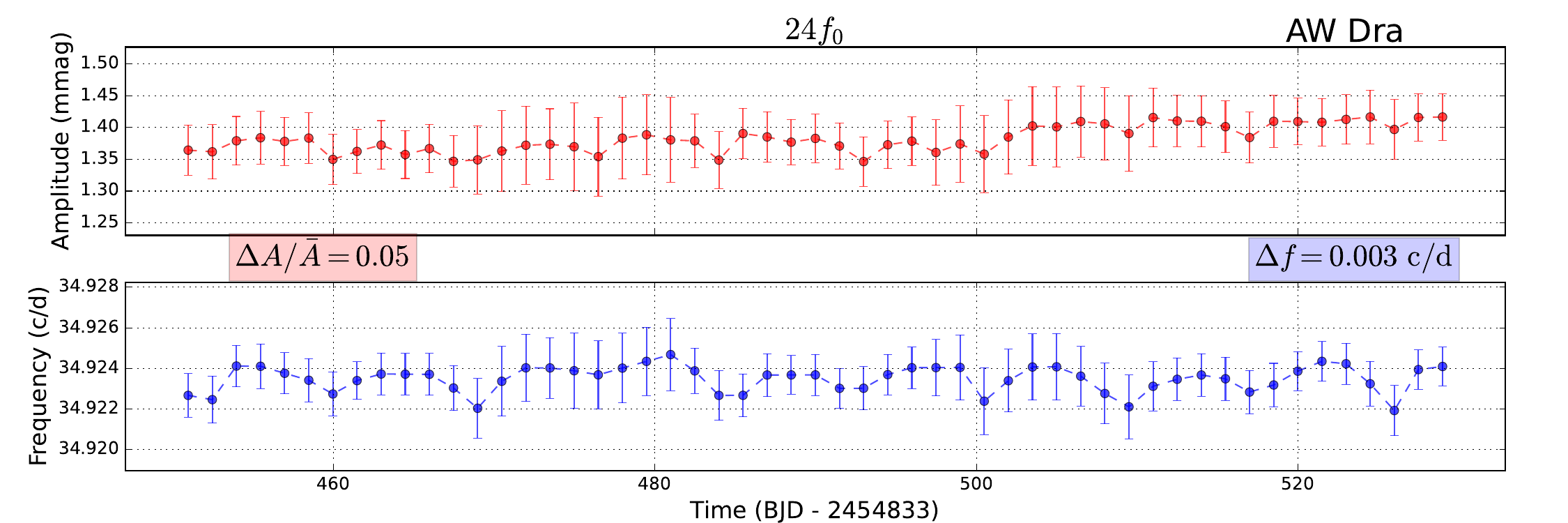}
  \includegraphics[width=0.48\textwidth]{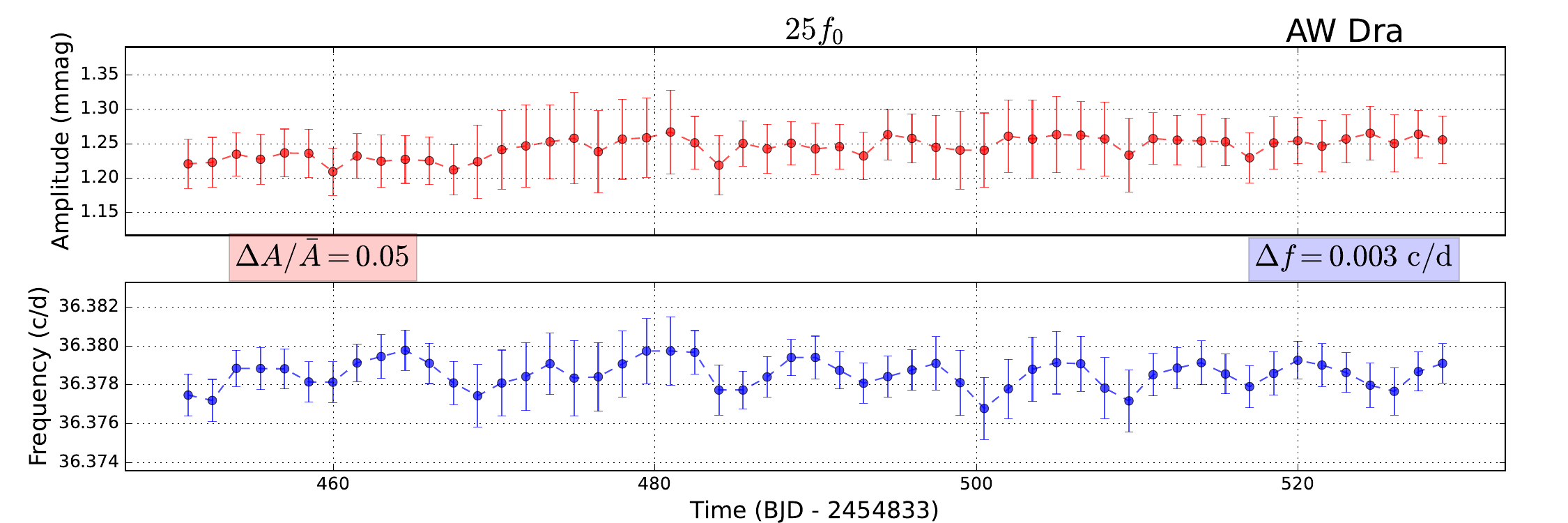}
  \includegraphics[width=0.48\textwidth]{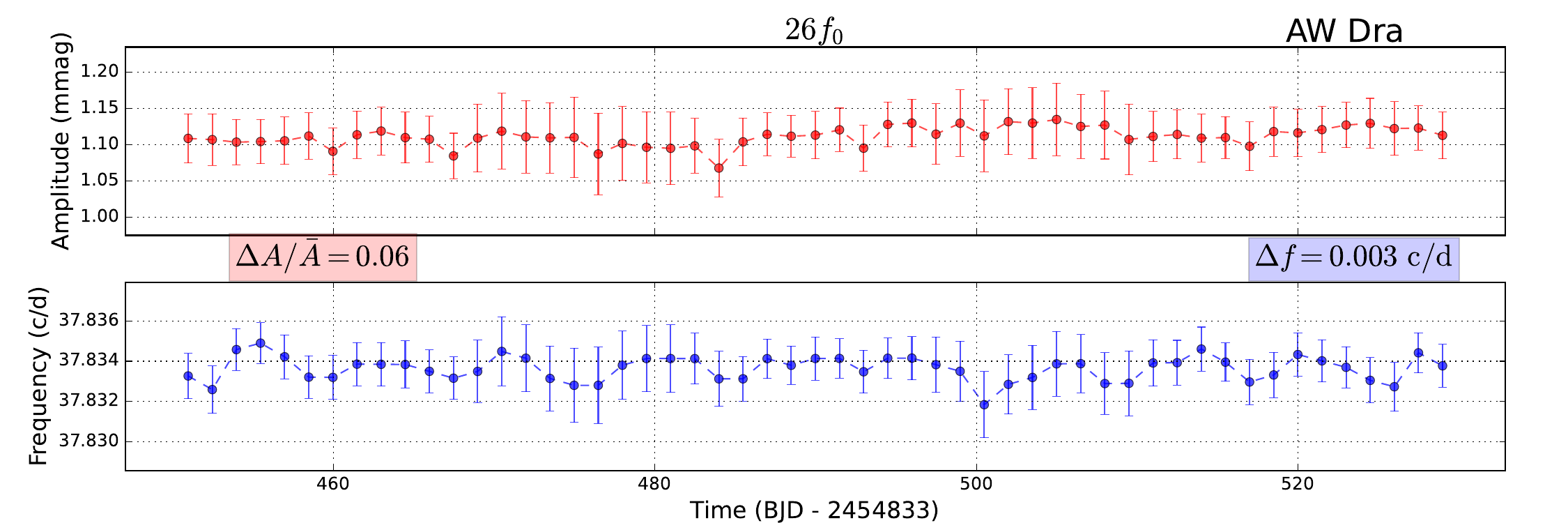}
  \includegraphics[width=0.48\textwidth]{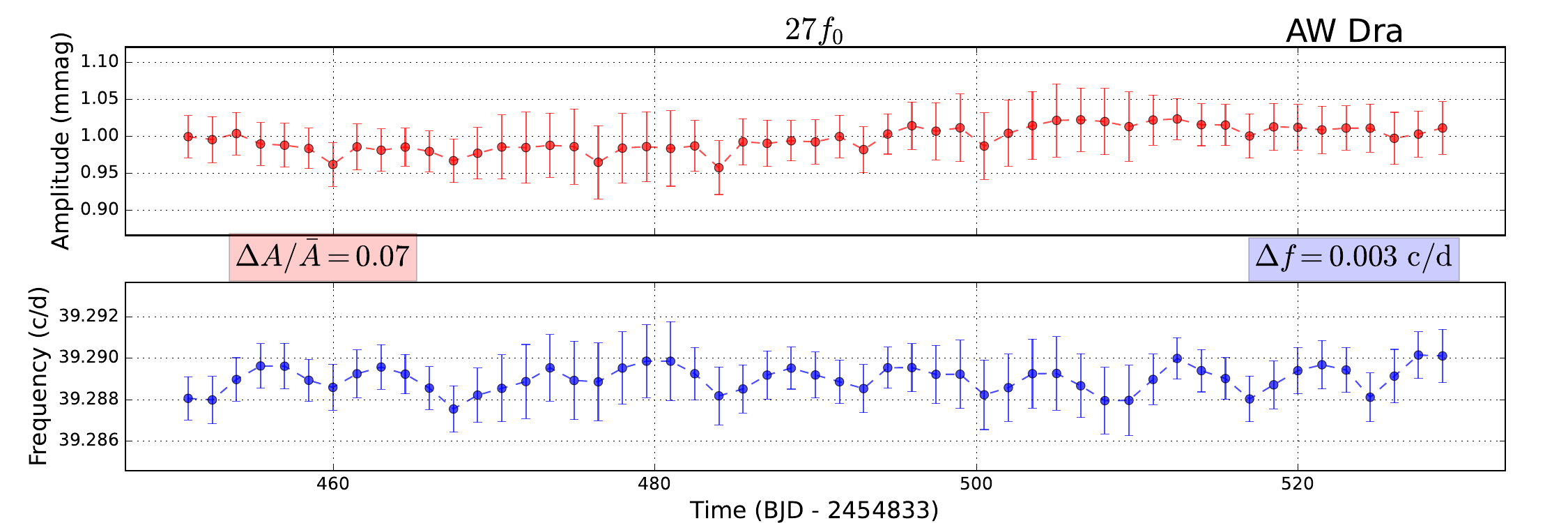}
  \includegraphics[width=0.48\textwidth]{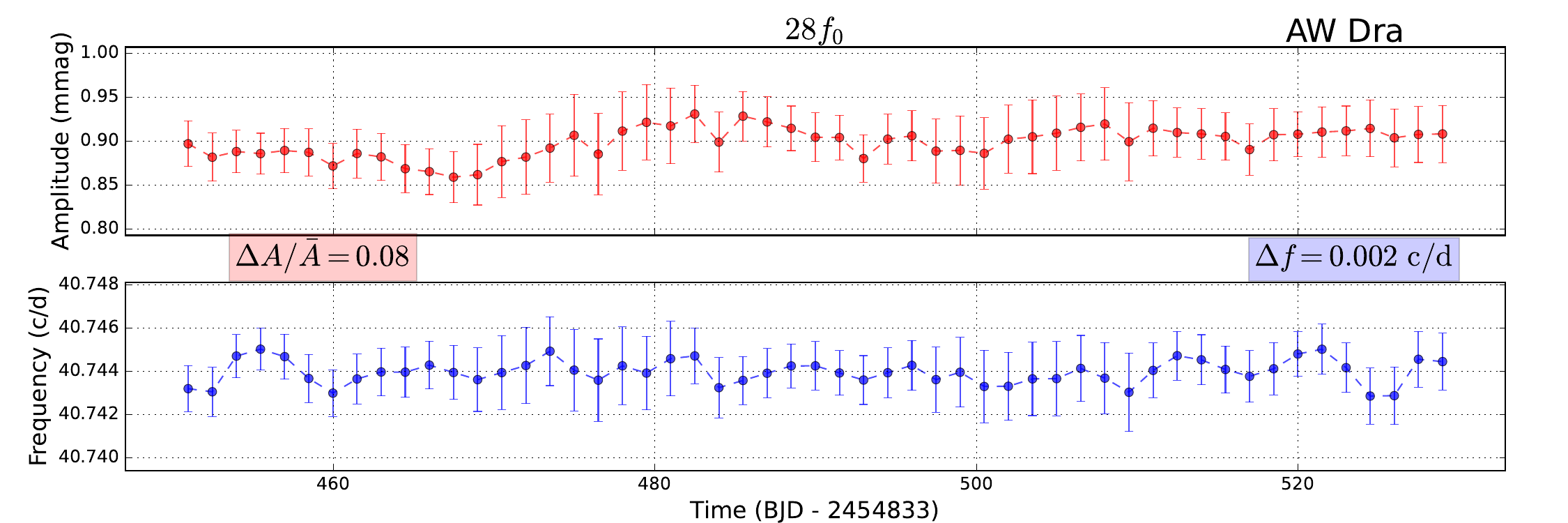}
  \includegraphics[width=0.48\textwidth]{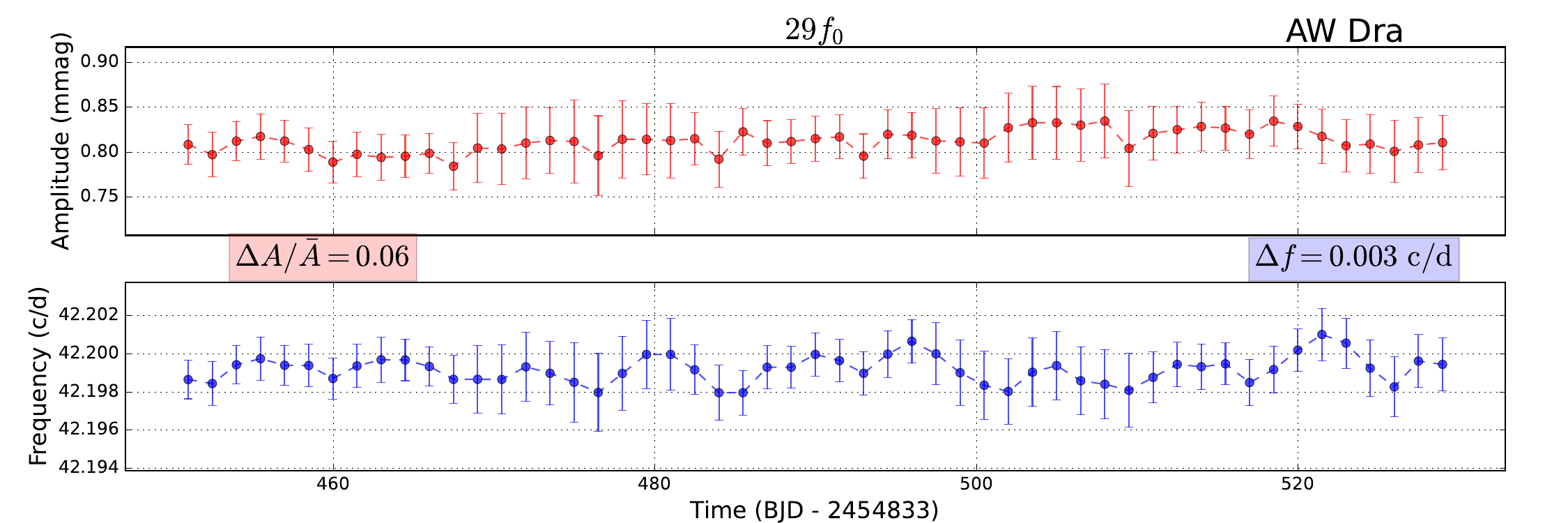}
  \includegraphics[width=0.48\textwidth]{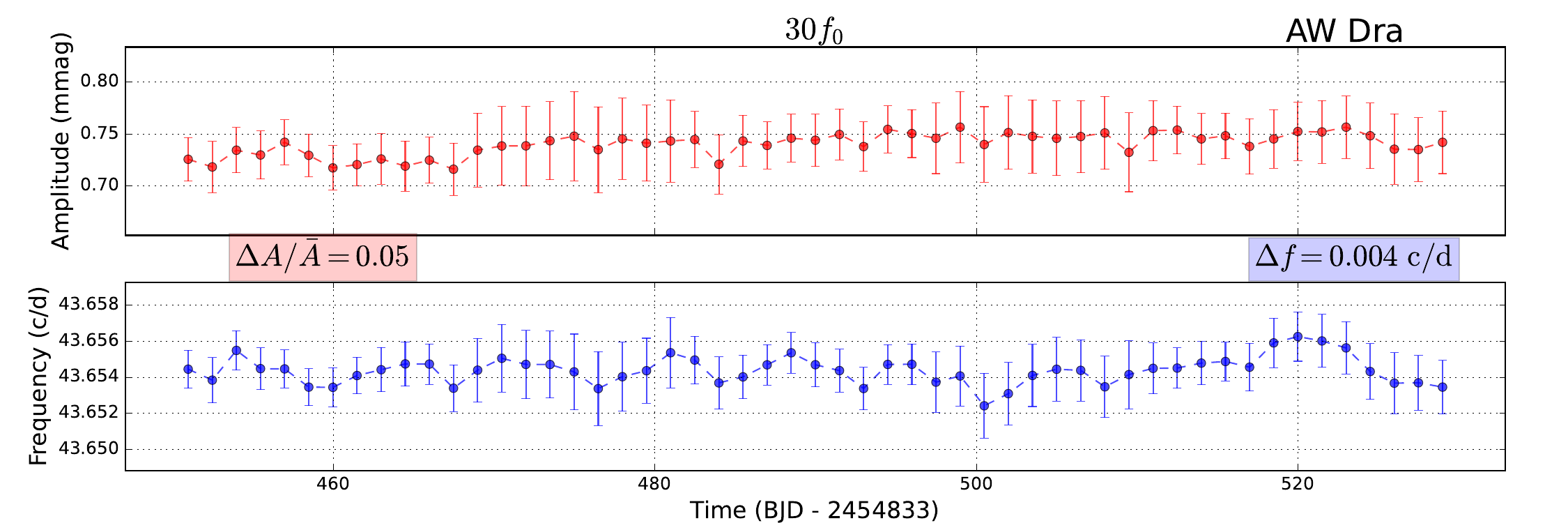}
  \includegraphics[width=0.48\textwidth]{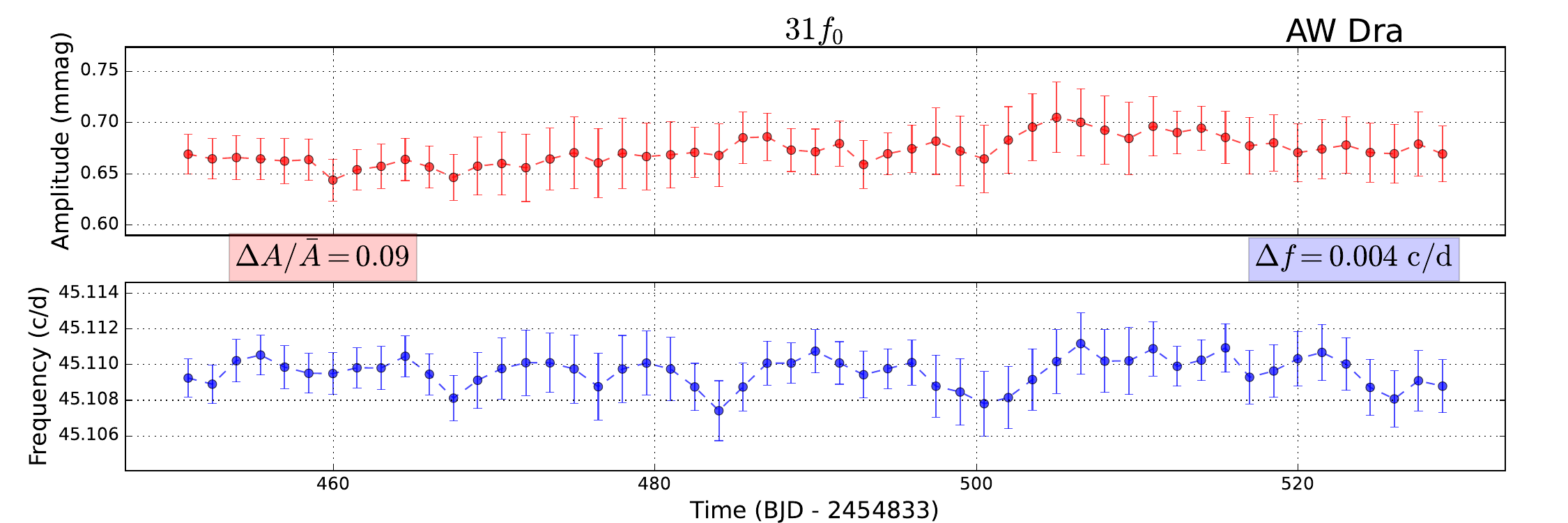}
  \includegraphics[width=0.48\textwidth]{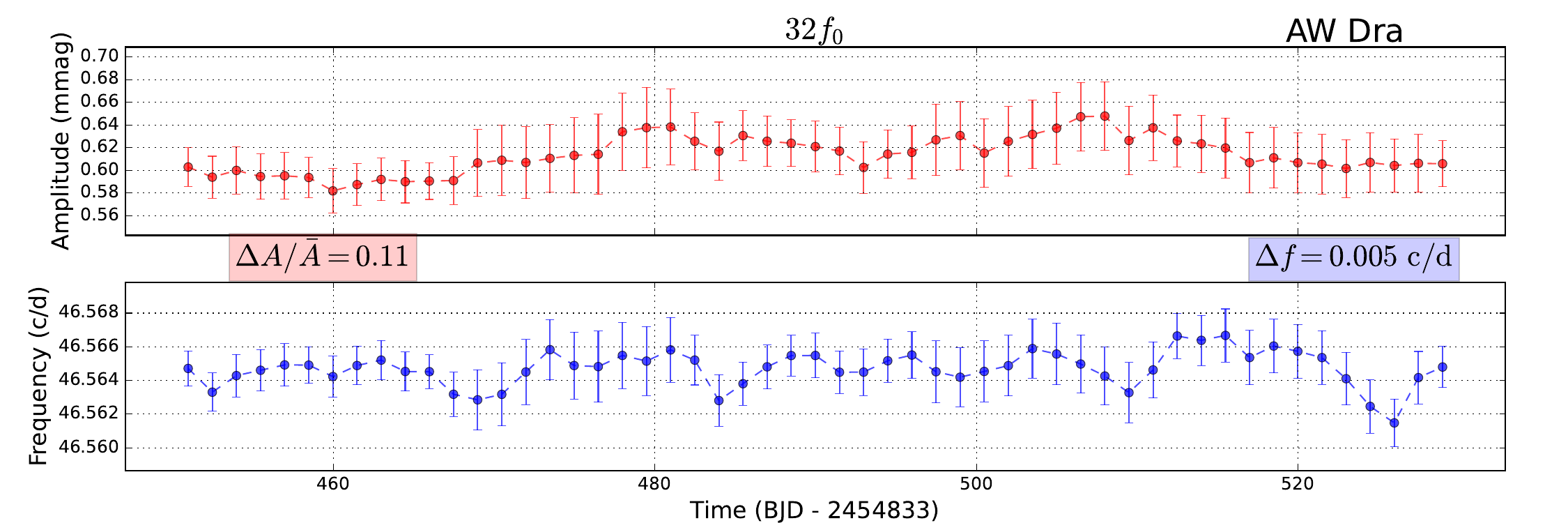}
  \includegraphics[width=0.48\textwidth]{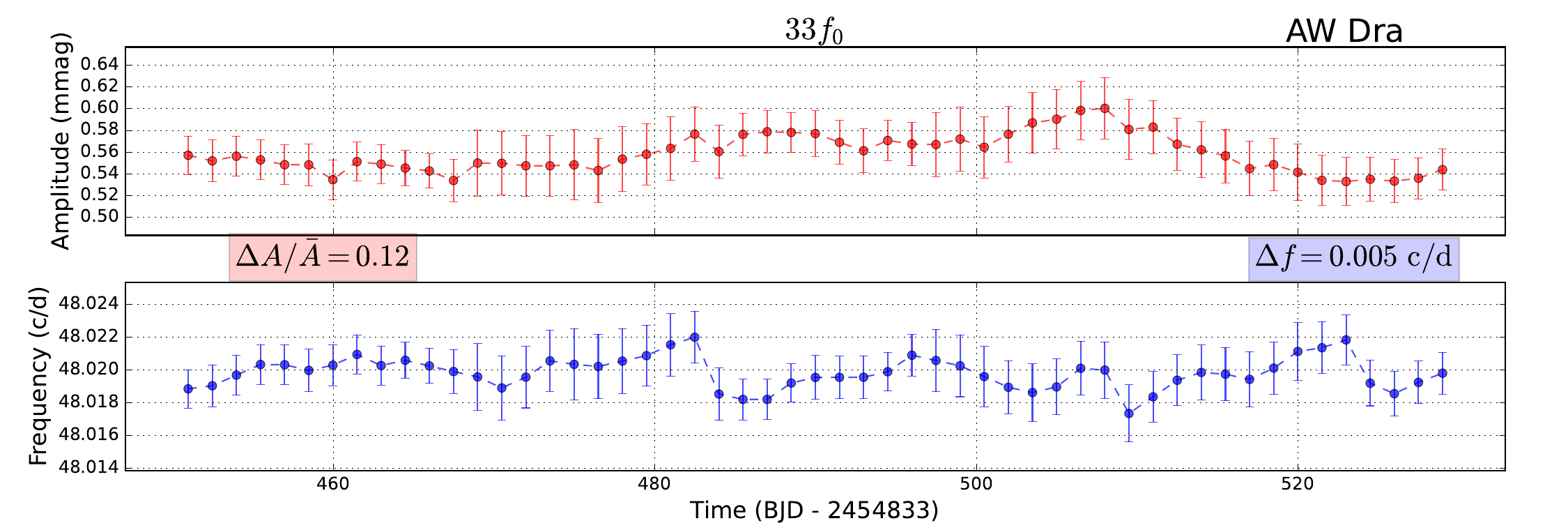}
  \includegraphics[width=0.48\textwidth]{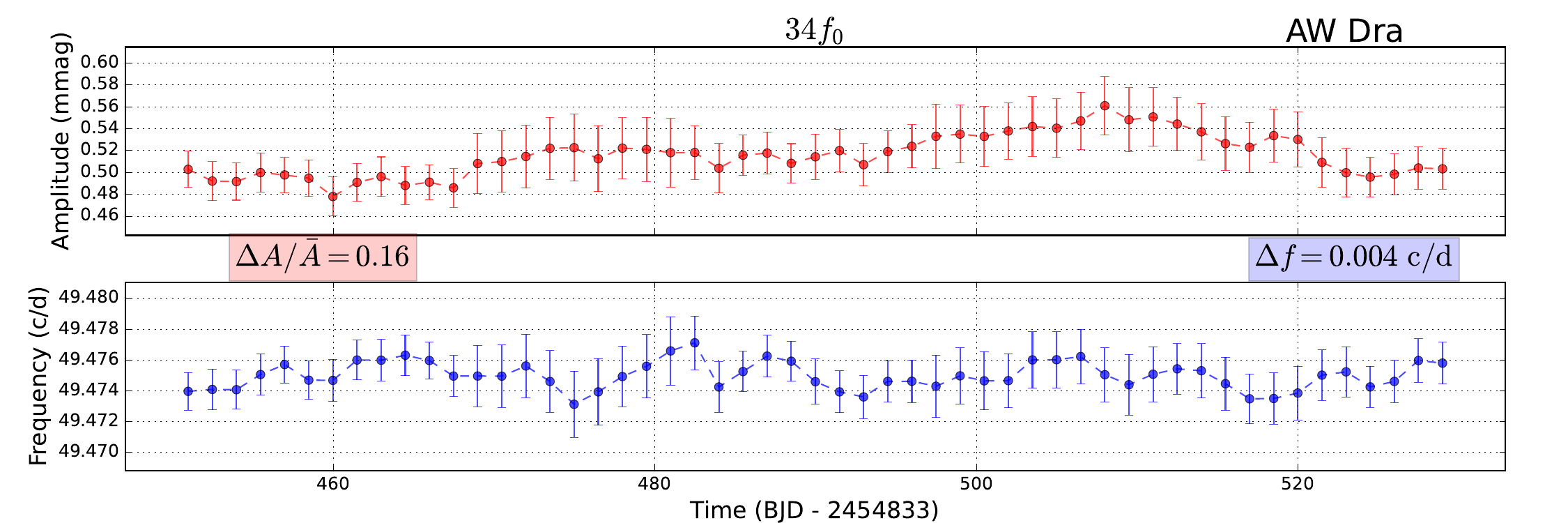}
  \includegraphics[width=0.48\textwidth]{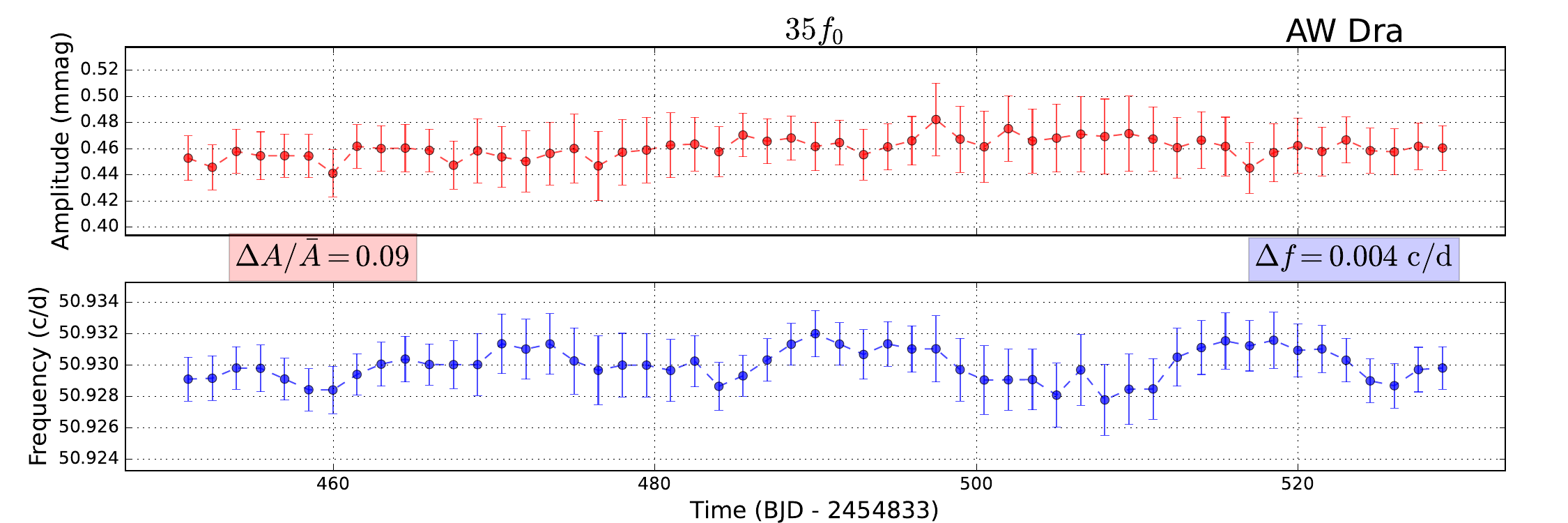}
  \includegraphics[width=0.48\textwidth]{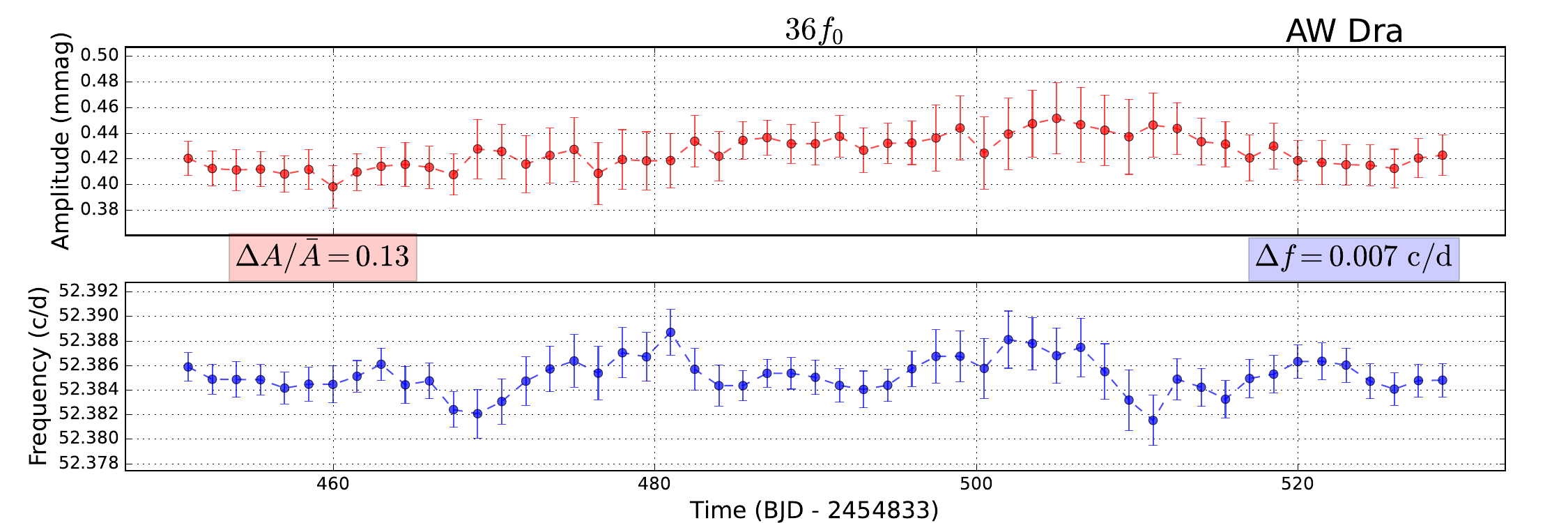}
  \caption{Temporal variations in amplitude and frequency for harmonics $f_0$--$36f_0$, part III.}
  \label{fig:var_amp_freq03}
\end{figure*}

\begin{table*}[hbtp!]
\centering
  \caption{Observational Parameters of the Detected Harmonics in AW Dra.}
  \label{tab:harmonics}
  \resizebox{0.7\textwidth}{!}{
  \begin{tabular}{l|ccccc}
    \hline
    \hline
  ID & Frequency (c/d) & Frequency Error (c/d) & Amplitude (mmag) & Amplitude Error (mmag) & S/N \\
    \hline
$f_0$	&  1.45516	& 0.00005 &304  	& 2   &128.3 \\
$2f_0$	&  2.91026	& 0.00004 &160  	& 1   &145.2 \\
$3f_0$	&  4.36541	& 0.00003 &105.5  	& 0.6   &167.1 \\
$4f_0$	&  5.82056	& 0.00004 & 52.1  	& 0.4   &143.3 \\
$5f_0$	&  7.27576	& 0.00004 & 32.7  	& 0.2   &150.7 \\
$6f_0$	&  8.73086	& 0.00003 & 20.3  	& 0.1   &173.9 \\
$7f_0$	& 10.18596  & 0.00004 &  8.06 	& 0.06  &135.5 \\
$8f_0$	& 11.64105  & 0.00004 &  5.13 	& 0.04  &140.5 \\
$9f_0$	& 13.09594  & 0.00005 &  2.33 	& 0.02  &106.2 \\
$10f_0$	& 14.55130  & 0.00005 &  1.92 	& 0.02  &120.9 \\
$11f_0$	& 16.00651  & 0.00005 &  2.93 	& 0.03  &116.4 \\
$12f_0$	& 17.46171  & 0.00005 &  3.34 	& 0.03  &107.2 \\
$13f_0$	& 18.91686  & 0.00006 &  3.41 	& 0.03  & 97.8 \\
$14f_0$	& 20.37201  & 0.00008 &  3.47 	& 0.05  & 74.4 \\
$15f_0$	& 21.82716  & 0.00006 &  3.29 	& 0.03  &105.8 \\
$16f_0$	& 23.28231  & 0.00005 &  3.13 	& 0.03  &107.0 \\
$17f_0$	& 24.73746  & 0.00006 &  2.91 	& 0.03  &105.1 \\
$18f_0$	& 26.19256  & 0.00006 &  2.65 	& 0.03  &101.5 \\
$19f_0$	& 27.64771  & 0.00006 &  2.41 	& 0.02  & 99.1 \\
$20f_0$	& 29.10286  & 0.00006 &  2.15 	& 0.02  & 98.4 \\
$21f_0$	& 30.55801  & 0.00006 &  1.94 	& 0.02  & 91.8 \\
$22f_0$	& 32.01317  & 0.00006 &  1.75 	& 0.02  & 95.3 \\
$23f_0$	& 33.46832  & 0.00007 &  1.56 	& 0.02  & 89.7 \\
$24f_0$	& 34.92347  & 0.00005 &  1.38 	& 0.01  &109.6 \\
$25f_0$	& 36.37857  & 0.00006 &  1.24 	& 0.01  &103.4 \\
$26f_0$	& 37.83372  & 0.00006 &  1.11 	& 0.01  &101.3 \\
$27f_0$	& 39.28887  & 0.00006 &  1.00 	& 0.01  & 98.9 \\
$28f_0$	& 40.74402  & 0.00006 &  0.90	& 0.01  & 94.4 \\
$29f_0$	& 42.19917  & 0.00007 &  0.810	& 0.009 & 88.8 \\
$30f_0$	& 43.65442  & 0.00007 &  0.738	& 0.009 & 85.4 \\
$31f_0$	& 45.10958  & 0.00007 &  0.670	& 0.008 & 80.3 \\
$32f_0$	& 46.56467  & 0.00008 &  0.608	& 0.008 & 77.4 \\
$33f_0$	& 48.01977  & 0.00007 &  0.558	& 0.007 & 78.5 \\
$34f_0$	& 49.47497  & 0.00007 &  0.513	& 0.006 & 88.5 \\
$35f_0$	& 50.93012  & 0.00008 &  0.463	& 0.006 & 72.1	\\
$36f_0$	& 52.38532  & 0.00008 &  0.425	& 0.006 & 69.1 \\
$37f_0$	& 53.84042  & 0.00009 &  0.388	& 0.006 & 63.5 \\
$38f_0$	& 55.2956   &  0.0001 &  0.347	& 0.006 & 59.3 \\
$39f_0$	& 56.7508   &  0.0001 &  0.322	& 0.006 & 55.9 \\
$40f_0$	& 58.2059   &  0.0001 &  0.293	& 0.005 & 53.6 \\
$41f_0$	& 59.6612   &  0.0001 &  0.266	& 0.005 & 52.2 \\
$42f_0$	& 61.1162   &  0.0001 &  0.247	& 0.005 & 47.2 \\
$43f_0$	& 62.5712   &  0.0001 &  0.226	& 0.005 & 44.5 \\
$44f_0$	& 64.0267   &  0.0001 &  0.211	& 0.005 & 44.5 \\
$45f_0$	& 65.4817   &  0.0001 &  0.200	& 0.005 & 42.3 \\
$46f_0$	& 66.9366   &  0.0001 &  0.172	& 0.004 & 39.4 \\
$47f_0$	& 68.3920   &  0.0002 &  0.160	& 0.004 & 36.1 \\
$48f_0$	& 69.8467   &  0.0002 &  0.150	& 0.004 & 35.1 \\
$49f_0$	& 71.3023   &  0.0002 &  0.140	& 0.004 & 33.9 \\
$50f_0$	& 72.7573   &  0.0002 &  0.124	& 0.004 & 29.3 \\
$51f_0$	& 74.2128   &  0.0002 &  0.114	& 0.004 & 27.1 \\
$52f_0$	& 75.6679   &  0.0002 &  0.107	& 0.004 & 26.7 \\
$53f_0$	& 77.1228   &  0.0002 &  0.099 & 0.004 & 25.7 \\
$54f_0$	& 78.5777   &  0.0003 &  0.087 & 0.004 & 22.1 \\
$55f_0$	& 80.0332   &  0.0003 &  0.085 & 0.004 & 21.4 \\
$56f_0$	& 81.4884   &  0.0003 &  0.081 & 0.004 & 20.5 \\
$57f_0$	& 82.9436   &  0.0003 &  0.074 & 0.004 & 19.1 \\
$58f_0$	& 84.3988   &  0.0003 &  0.067 & 0.004 & 17.4 \\
$59f_0$	& 85.8538   &  0.0004 &  0.062 & 0.004 & 16.5 \\
$60f_0$	& 87.3083   &  0.0004 &  0.059 & 0.004 & 15.7 \\
$61f_0$	& 88.7641   &  0.0004 &  0.048 & 0.004 & 13.5 \\
$62f_0$	& 90.2186   &  0.0004 &  0.050 & 0.004 & 13.2 \\
$63f_0$	& 91.6729   &  0.0005 &  0.045 & 0.004 & 12.0 \\
$64f_0$	& 93.1292   &  0.0005 &  0.042 & 0.004 & 11.8 \\
$65f_0$	& 94.5841   &  0.0005 &  0.038 & 0.003 & 11.0 \\
$66f_0$	& 96.0385   &  0.0006 &  0.037 & 0.004 & 10.4 \\
$67f_0$	& 97.4952   &  0.0006 &  0.038 & 0.004 & 10.2 \\
$68f_0$	& 98.9492   &  0.0007 &  0.029 & 0.004 &  8.2 \\
$69f_0$	&100.4046   &  0.0007 &  0.030 & 0.003 &  8.7 \\
$70f_0$	&101.8600   &  0.0007 &  0.027 & 0.003 &  8.5 \\
\hline
\end{tabular}
}
\end{table*}

\end{appendix}
\end{document}